\documentclass[11pt,fleqn]{article} 
\usepackage{graphicx}
\usepackage{textcomp} 
\usepackage[pdftex]{pict2e}
\usepackage{epstopdf}
\usepackage{latexsym}
\usepackage{amsmath,mathrsfs}
\usepackage{amssymb}
\usepackage{bm}
\usepackage[mathcal]{euscript}
\usepackage{slashed}
\usepackage{mathbbol}

\usepackage{genyoungtabtikz}
\Yboxdim{5pt}
\Ylinethick{0.8pt}
\newcommand{\yngRed}[1]{%
    {\Ylinecolour{red}
    {\gyoung(#1)}}%
}
\newcommand{\yngBlue}[1]{%
    {\Ylinecolour{blue}\gyoung(#1)}%
}

\newcommand{\bluefill}{\Yfillcolour{blue}}

\newcommand{\shiftleft}[2]{\makebox[0pt][r]{\makebox[#1][l]{#2}}}

\usepackage[top=30mm, bottom=30mm, left=25mm, right=25mm]{geometry}

\numberwithin{equation}{section}

\usepackage{marvosym}

\newcommand\blfootnote[1]{
  \begingroup
  \renewcommand\thefootnote{}\footnote{#1}
  \addtocounter{footnote}{-1}
  \endgroup
}

\usepackage{hyperref}
\hypersetup{
        unicode,	
        colorlinks,
        citecolor=blue,
        linkcolor=blue, 
        urlcolor=red,
        bookmarksopen=true,
        bookmarksopenlevel=\maxdimen,
      }

\def\gl#1#2{\ifmmode \mathrm{GL}(#1; {\bf #2}) \else $\mathrm{GL}(#1; {\bf #2})$\fi}
\def\sl#1#2{\ifmmode \mathrm{SL}(#1; {\bf #2}) \else $\mathrm{SL}(#1; {\bf #2})$\fi}
\def\so#1{\ifmmode \mathrm{SO}({#1}) \else $\mathrm{SO}(#1)$\fi}

\def\sp#1#2{\ifmmode \mathrm{Sp}(#1; {\bf #2}) \else $\mathrm{Sp}(#1; {\bf #2})$\fi}
\def\usp#1{\ifmmode \mathrm{USp}(#1) \else $\mathrm{USp}(#1)$\fi}
\def\spin#1{\ifmmode \mathrm{Spin}(#1) \else $\mathrm{Spin}(#1)$\fi}
\def\su#1{\ifmmode \mathrm{SU}({#1}) \else $\mathrm{SU}(#1)$\fi}

\def\double #1{#1{\hbox{\kern-2pt $#1$}}}

\mathcode`\*="702A                  
\def\half{{\textstyle{1\over{\raise.1ex\hbox{$\scriptstyle{2}$}}}}}

\def \a{\alpha}

\def \b{\beta}

\def \d{\delta}

\def \O{\Omega}

\def\s{\sigma}

\def\wdg{\wedge}
\def\Comp{\mathbb{C}}
\def\blueBullet{\textcolor{blue}{\bullet}}
\def\redBullet{\textcolor{red}{\bullet}}

\begin{document}

\begin{flushright}
\makebox[0pt][b]{}
\end{flushright}

\vspace{10pt}
\begin{center}
{\LARGE Unfolding the Six-Dimensional Tensor Multiplet}

\vspace{40pt}
Carlo Iazeolla${}^{\diamondsuit,\ast}$, Per Sundell${}^{\clubsuit
}$, Brenno Carlini Vallilo${}^{\spadesuit}$
\vspace{30pt}

{\em Addresses\\ 
\vspace{5pt}
${}^\diamondsuit$ Dipartimento di Scienze Ingegneristiche,  G. Marconi University \\  Via Plinio 44, 00193, Roma, Italy \& \\Sezione INFN Roma “Tor Vergata” \\ Via della Ricerca Scientifica 1, 00133, Roma, Italy\\
\vspace{5pt}
${}^{\clubsuit}$  Instituto de Ciencias Exactas y Naturales, Universidad Arturo Prat,\\ Playa Brava 3265, 1111346 Iquique, Chile \& \\
Facultad de Ciencias, Universidad Arturo Prat,\\ Avenida Arturo Prat Chacón 2120, 1110939 Iquique, Chile\\
\vspace{5pt}
${}^{\spadesuit}$ Departamento de Física y Astronimía, Facultad de Ciencias Exactas,\\ Universidad Andres Bello, 
Sazié 2212, Santiago, Chile\\
\vspace{5pt}
}

\vspace{40pt}
{\bf Abstract}
\end{center}
We derive a manifestly superconformally covariant unfolded formulation of the free $(2,0)$ tensor multiplet in six spacetime dimensions.  
The unfolded system consists of an abelian two-form and an infinite-dimensional chiral zero-form containing the system's curvature, matter fields, and their derivatives on-shell, realized using superoscillators.
The construction of the cocycle gluing these forms on a general superconformal background goes one step beyond previous results in super-Poincaré backgrounds.

\blfootnote{
${}^\diamondsuit$ \href{mailto:}{c.iazeolla@gmail.com}, 
${}^{\clubsuit}$ \href{mailto:per.anders.sundell@gmail.com}{per.anders.sundell@gmail.com}, 
${}^{\spadesuit}$ \href{mailto:vallilo@unab.cl}{vallilo@unab.cl} }
\blfootnote{${}^\ast$ Member of INDAM-GNFM.}

\setcounter{page}0
\thispagestyle{empty}

\newpage

\tableofcontents

\parskip = 0.1in

\section{Introduction and Motivation}

The second superstring revolution revealed a web of non-perturbative dualities among the known perturbative string theories, tied together by eleven-dimensional supergravity and cemented by overwhelming evidence about various strong coupling limits of said theories. 
In this web, extended supersymmetric objects play a key role in providing observables that can be computed on both sides of dualities, tied together into a superconformal web conjectured to be crowned by six-dimensional superconformal field theories related to five-branes \cite{Berkooz:1997cq, Seiberg:1997ax, Aharony:1997th}.

Three decades since their discovery, a direct, non-bootstrap formulation of these theories remains elusive. However, they are expected to exhibit $(2,0)$  supersymmetry, extending to the full six-dimensional superconformal group  $OSp(8^\ast|4)$.
As these theories lack a dimensionless parameter, it is widely accepted that, apart from the free case, they do not admit a Lagrangian formulation using actions built from integrating densities over six-dimensional spacetime. Indeed, such a formulation immediately stumbles upon the fact that the free, six-dimensional $(2,0)$ multiplet includes an anti-selfdual two-form. 

Formulations of non-conformal, single five-branes \cite{Pasti:1995tn, Cederwall:1997gg, Howe:1996yn, Bansal:2023pnr} are available, which one may view as non-linear, Born-Infeld-like extensions of the free on-shell superspace description of the multiplet  \cite{Howe:1983fr, Koller:1982cs}. Additionally, five-branes are expected to contain extended objects, such as selfdual strings arising from boundaries of open membranes. However, in the case of conformal, coinciding stacks of five-branes, these extended objects do not possess a characteristic size as expected for tensionless objects. It can be argued that without extended objects the theory is free, e.g., \cite{DouglasTalk}. 
Indeed, considering stacks of coinciding five-branes, it is reasonable to expect infinite spectra of local degrees of freedom in six spacetime dimensions carried by real, chiral field strengths of ever-increasing spins that cannot be consistently truncated without reducing to a free theory. An effective interacting theory is necessarily non-local since there is no dimensional constant in which a local perturbative expansion can be computed, as in the case of ten-dimensional string theory. 

The stagnation of conventional QFT approaches to construct interacting, six-dimensional, conformal field theories suggests the need for an alternative perspective. Indeed, the root cause of the challenges facing interacting (2,0) models parallel those of higher-spin gravity. The latter, however, has been provided with a nontrivial on-shell formulation using Vasiliev’s unfolded formalism \cite{Vasiliev:1988sa,Vasiliev90,properties,more,Vasiliev:1999ba,Bekaert:2004qos,Didenko:2014dwa}.
While the Lagrangian formulation for the full higher-spin gravity directly in spacetime remains an open problem --- despite remarkable results obtained for complex, chiral models in four spacetime dimensions \cite{Ponomarev:2016lrm,Sharapov:2022faa,Sharapov:2022awp,Didenko:2022qga,Sharapov:2023erv} --- progress has instead been made \cite{BarnichGrigoriev,Boulanger:2011dd,Vasiliev:2011knf, Boulanger:2015kfa,Bonezzi:2016ttk,FSG1,FSG2} towards an off-shell formulation of Vasiliev's theory as an Alexandrov-Kontsevich-Schwarz-Zaboronsky (AKSZ) sigma model \cite{AKSZ}.

The AKSZ formalism uses graded geometry tools to solve the BV quantum master equation (without the need for any perturbative measure corrections), thus providing an analog in the context of partial differential equations to the Hamiltonian formalism for ordinary differential equations. 
Although developed initially mainly within the context of lower-dimensional topological field theories, where it played a central role, for example, in the deformation quantization of Poisson geometries \cite{Kontsevich,CaFe}, the AKSZ formalism also applies to relativistic field theories with local degrees of freedom encoded into infinite-dimensional target geometries.
To this end, the original equations of motion are extended, or unfolded, into universally Cartan integrable systems of constraints on generalized curvatures built from differential forms.
These systems are then treated as boundary conditions to topological field theories extended into bulk manifolds with action functionals facilitating the construction of quantization functors mapping the original classical solution spaces into boundary operator algebras \cite{FSG1,FSG2}.

The resulting formalism has two remarkable features: 

\noindent a) Besides being manifestly diffeomorphism covariant, it can be made compatible with the larger group of similarity transformations of noncommutative  geometries, referred to as Kontsevich gauge transformation, provided the internal symmetry algebras are taken to be homotopy associative instead of homotopy Lie \cite{Boulanger:2011dd,Sharapov:2017yde,meta,FSG1}; and 

\noindent b) It geometrically quantizes the targets of the AKSZ sigma models, which means that it applies naturally to chiral theories provided their targets can be equipped with nontrivial Poisson structures.

Combining the unfolded approach to partial differential equations with the AKSZ approach to quantization yields a more general framework for gauge theory than the metric-like approach. Any set of local, diffeomorphism-covariant partial differential equations on a metric space can be unfolded, while the converse does not generally hold.

In this paper, we unfold the linearized, six-dimensional, superconformal, $(2,0)$ tensor multiplet, which contains an anti-selfdual three-form curvature tensor.
Our main results, which extend those of 
\cite{Grojean:1998zt}, are:

\noindent i) the superconformal cocycle that glues the conformally primary zero-forms to the two-form potential; 

\noindent ii) the superspace constraints\footnote{From the local homotopy invariance of universally Cartan integrable systems, it follows that the superspace and component formulations of a supersymmetric field theory share the same unfolded formulation; for examples, see \cite{Engquist:2002gy,Grigoriev:2025vsl}.} on the superconformal \emph{chiral zero-form module}\footnote{
The local degrees of freedom of an unfolded relativistic field theory, i.e., its Cartan integration constants, constitute an infinite-dimensional module of the symmetry algebra in form degree zero. 
The module can be given in various bases; in its Lorentz covariant basis, the primary tensors correspond to on-shell curvatures obeying Bargmann-Wigner equations, e.g., the Faraday and Weyl tensors in the case of unfolded electromagnetism and gravity, respectively \cite{Vasiliev:1999ba,Bekaert:2004qos,Didenko:2014dwa}.
Generally, the zero-form module consists of all on-shell curvatures tensors (including scalar fields) and their descendants.
We refer to the zero-form module of the (2,0) model as its chiral zero-form module.
}.

As we shall comment in the Conclusions and Prospects, our results pave the way towards unfolding the linearized version of Hull's $(4,0)$ model \cite{Hull:2000zn,Hull:2000rr}, and a doubled conformal superspace providing an alternative to \cite{Butter:2009cp}.
Our analysis also provides a starting point for quantizing $(2k,0)$ systems using AKSZ sigma models with targets built from Hermitian modules of the complexification of the metaplectic group ${\rm Mp}(16;\mathbb{R})$ by generalizing the formalism set up for ${\rm Mp}(4;\mathbb{R})$ in the context of three-dimensional conformal field theories dual to four-dimensional higher-spin gravities\footnote{The metaplectic group ${\rm Mp}(8;\mathbb{R})$ enters the construction of AKSZ sigma models for four-dimensional conformal field theories.}.

\paragraph{Organization of the paper.} In section \ref{superconnections}, we outline the general framework of unfolded dynamics and quantization. Section \ref{algebraicstructures} describes the superconformal algebraic structure of our formalism. In section \ref{background}, we define the vacuum connections associated with the global symmetries of the theory. Next, we define the two-form potential and derive the cocycle sourcing it in Section \ref{poincarebackground}. In section \ref{superconformalbackground}, we extend the result of the previous section for a flat superconformal background and thus derive the superconformal transformations for the field strengths. We conclude our work and point out further directions for research in Section \ref{conclusion}.
The appendix contains the notation used, $\sigma$-matrices conventions and identities, and a discussion on the symmetries of unfolded systems.

\section{Quillen Superconnections}
\label{superconnections}

This section describes how differential forms valued in graded algebras arise naturally as fundamental fields of AKSZ sigma models suitable for including chiral bosons and noncommutative  geometries into a quantum field theory framework that we expect to play a role in $(2,0)$ theories.

\paragraph{Basic structures.}

The basic ingredient is as a differential, graded, Frobenius algebra, i.e., an operator algebra $\boldsymbol{\mathcal{V}}$ 
equipped with i) a $\mathbb{Z}$-degree map; ii) a degree-preserving, binary, associative product represented in what follows by juxtaposition; iii) a nilpotent differential $D$ of degree one compatible with (ii); and iv) a graded-cyclic, non-degenerate, real-valued trace operation ${\rm str}_{\boldsymbol{\mathcal{V}}}$ of degree $-3$ compatible with (i)--(iii).
Requiring (i)--(iii) yields a differential, graded, associative algebra (DGA).
Taking the dynamical variable to be an element $\boldsymbol{z}\in \boldsymbol{\mathcal{V}}$ of degree $+1$, alias, a Quillen superconnection, the action functional 
\begin{align}\label{S}
S={\rm str}_{\boldsymbol{\mathcal{V}}}\left(\frac12 \boldsymbol{z}D\boldsymbol{z}+\frac13 \boldsymbol{z}^3\right)
\end{align}
has stationary points given by flat superconnections, viz.,
\begin{align}
D\boldsymbol{z}+\boldsymbol{z}^2\approx 0\ .
\end{align} 
A class of noncommutative   sigma models arise by taking $\boldsymbol{\mathcal{V}}=\Omega^{\mathcal{S}}(\boldsymbol{M})\otimes\boldsymbol{\mathcal{A}}$, where $\Omega^{\mathcal{S}}(\boldsymbol{M})$ is a Frobenius DGA represented by a set $\mathcal{S}$ of differential forms on a noncommutative   differential Poisson manifold $\boldsymbol{M}$, referred to as the symbols of $\Omega^{\mathcal{S}}(\boldsymbol{M})$, and an internal Frobenius DGA $\boldsymbol{\mathcal{A}}$ with differential $\delta$, such that $D = d + \delta$, and ${\rm str}_{\boldsymbol{\mathcal{V}}}=\int_{\boldsymbol{M}} {\rm str}_{\boldsymbol{\mathcal{A}}}$, where $\int_{\boldsymbol{M}}$ is the Cartan measure on $\boldsymbol{M}$ of degree $-{\rm dim}(\boldsymbol{M})$, and ${\rm str}_{\boldsymbol{\mathcal{A}}}$ has degree ${\rm dim}(\boldsymbol{M})-3$. 
The superconnection $\boldsymbol{z}$ thus consists of differential forms on $\boldsymbol{M}$ valued in $\boldsymbol{\mathcal{A}}$; in particular, the one-forms gauge the associative subalgebra of $\boldsymbol{\mathcal{A}}$ in degree zero.
Taking ${\rm dim}(\boldsymbol{M})=3$, and $\boldsymbol{\mathcal{A}}=\boldsymbol{\mathcal{K}}$, an associative Frobenius algebra with vanishing degree map, $\delta=0$, and ${\rm str}_{\boldsymbol{\mathcal{A}}}={\rm tr}_{\boldsymbol{\mathcal{K}}}$, yields noncommutative   Chern--Simons theory, reducing to ordinary Chern--Simons theory in the commutative case, for which $\boldsymbol{\mathcal{K}}$ can be restricted further to Lie subalgebras.
Taking instead ${\rm dim}(\boldsymbol{M})=4$, $\boldsymbol{\mathcal{A}}=\boldsymbol{\mathcal{K}} \oplus \rho_{[-1]}\otimes K$, where $\boldsymbol{\mathcal{K}}$ is now unital and $\rho_{[-1]}$ is a nilpotent generator of degree $-1$, and $\delta \rho_{[-1]}=1$ and ${\rm str}_{\boldsymbol{\mathcal{A}}} \, (x+\rho_{[-1]} y)={\rm tr}_{\boldsymbol{\mathcal{K}}}\, y$, and decomposing $\boldsymbol{z}=A+\rho_{[-1]} B$, yields the  action of noncommutative   BF theory, viz.,
\begin{align}
S=\int_{\boldsymbol{M}} {\rm tr}_{\boldsymbol{\mathcal{K}}}\,\left(B (dA+A A)+\frac12 B B\right)\ .
\end{align}
Models with local degrees of freedom can be achieved by taking $\boldsymbol{\mathcal{A}}$ to contain infinite-dimensional subspaces in degree zero, e.g., Frobenius--Chern--Simons models containing higher-spin gravities of Vasiliev type \cite{Boulanger:2015kfa,Bonezzi:2016ttk} and three-dimensional conformal field theories \cite{FSG1,FSG2} as classical moduli spaces.
In the absence of zero-form constraints, the classical action \eqref{S} can be completed into a BV master action by extending the Quillen superconnection into an AKSZ superconnection, that we shall denote by $\boldsymbol{z}$ as well, containing additional fields and anti-fields of non-trivial ghost numbers, by introducing a total degree given by the sum of $\boldsymbol{\mathcal{V}}$-degree and ghost number and taking the AKSZ superconnection to have total degree $+1$.

\paragraph{BV geometry.}

AKSZ sigma models are particular instances of BV quantum field theories. The latter have off-shell configuration spaces that are BV manifolds $(\mathcal{M},\mu_{\mathcal M},\omega_{\mathcal M})$ equipped with divergence-free, Hamiltonian Q-structures, denoted by $\vec s$.
Thus, the off-shell field configurations of a BV model belong to a $\mathbb{Z}$-graded supermanifold $\mathcal{M}$ equipped with i) a Berezin measure $\mu_{\mathcal M}$ of degree zero inducing a divergence operation, viz., ${\rm div}(\vec v)\mu_{\mathcal M}:={\rm Lie}_{\vec v}\,u_{\mathcal M}$; and ii)  a symplectic structure $\omega_{\mathcal M}$ of degree $-1$ corresponding to a Poisson structure of degree $+1$ with bracket $\{\cdot,\cdot\}_{\mathcal M}$.
Together, (i) and (ii) induce a nilpotent Laplacian $\Delta:\Omega_{[0]}(\mathcal{M})\to \Omega_{[0]}(\mathcal{M})$ of degree one via $\Delta\phi:= \frac12 {\rm div}(\vec X_\phi)$, where $\vec X_\phi$ is the Hamiltonian vector field generated by $\phi\in \Omega_{[0]}(\mathcal{M})$; the degree map, measure, bracket, and Laplacian, are referred to, respectively, as the ghost number, S-structure, BV bracket, or P-structure, and BV Laplacian.
The divergence-free, Hamiltonian Q-structure 
\begin{align}
\vec s :=\vec X_S\ ,
\end{align}
for an $S\in \Omega_{[0]}(\mathcal{M})$ of degree zero obeying
\begin{align}
\{S,S\}_{\mathcal M}=0\ ,\qquad \Delta S=0\ ,
\end{align}
referred to as a master action.
The Lagrangian submanifolds $\mathcal{L}\subset \mathcal{M}$ define gauge fixings; the vanishing locus $\mathcal{C}$ of $\vec s$ defines the extended classical solution space; and the intersections $\mathcal{L}\cap\mathcal{C}$ define gauge-fixed classical solution spaces. 
$\mathcal{C}$ decomposes into gauge orbits generated by trivial vector fields $\vec t=[\vec s,\vec u]$ generated by (arbitrary) vectors fields $\vec u$ of ghost number $-1$.
To avoid trivialities, the space of classical observables is defined as the restriction of the $\vec s$-cohomology to $\mathcal{C}$, and the space of quantum observables as the restriction of the $\vec s$-cohomology to the kernel of $\Delta$.
Non-trivially, the measure on $\mathcal{M}$ induces measures $\mu_{\mathcal{L}}$ on $\mathcal{L}$, such that quantum observables $O$ can be assigned expectation $\int_{\mathcal{L}} \mu_{\mathcal L} e^{iS} O$ depending $\mathcal{L}$ via its homotopy class in $\mathcal{M}$.
The conditions on $\Delta S$ and $\Delta O$ can be relaxed to the quantum master equations, albeit at the expense of introducing singular quantities. 

\paragraph{AKSZ sigma model.}
BV geometries can be induced geometrically by taking $\mathcal M$ to consist of maps from a set of Q-manifolds
$\{(\boldsymbol{X},\vec{Q}_{\boldsymbol{X}},\mu_{\boldsymbol{X}})\}$, referred to as the sources, equipped with compatible Berezin measures $\mu_{\boldsymbol{X}}$, i.e., $\int_{\boldsymbol{X}}\mu_{\boldsymbol{X}} \vec  Q_{\boldsymbol{X}}(\cdot)=0$, into a fixed Q-manifold $(\boldsymbol{Y},\vec{Q}_{\boldsymbol{Y}},\{\omega_{\boldsymbol{Y}},H_{\boldsymbol{Y}}\})$, referred to as the target, equipped with a set of compatible graded symplectic structures $\omega_{\boldsymbol{Y}}$, viz.,
\begin{align}
\imath_{\vec{Q}_{\boldsymbol{Y}}}\omega_{\boldsymbol{Y}}=dH_{\boldsymbol{Y}}\ ,\qquad {\rm deg}_{\boldsymbol{Y}}(\omega_{\boldsymbol{Y}})={\rm deg}_{\boldsymbol{Y}}(H_{\boldsymbol{Y}})-1\ ,
\end{align}
for bracket-nilpotent Hamiltonian functions $H_{\boldsymbol{Y}}$.
Thus, 
\begin{align}
\mathcal{M}=\bigcup_{\boldsymbol{X}} \mathcal{M}(\boldsymbol{X})\ ,\qquad \mathcal{M}(\boldsymbol{X})=\{\varphi:\boldsymbol{X}\to \boldsymbol{Y}\}\ ,
\end{align}
such that ${\rm Vec}(\mathcal{M})=\bigoplus_{\boldsymbol{X}} {\rm Vec}(\mathcal{M}(\boldsymbol{X}))$.
The left- and right-actions of ${\rm Diff}(\boldsymbol{Y})$ and ${\rm Diff}(\boldsymbol{X})$ on ${\mathcal M}_{\boldsymbol{X}}$ induce Lie-algebra monomorphisms 
\begin{align}
(\cdot)_{\mathcal{M}}: {\rm Vec}(\boldsymbol{Y})\to  {\rm Vec}(\mathcal M)
\ ,\qquad (\cdot)_{\mathcal{M}(\boldsymbol{X})}: {\rm Vec}(\boldsymbol{X})\to  {\rm Vec}(\mathcal M(\boldsymbol{X}))\ ,
\end{align}
whose images are mutually graded-commuting subalgebras.
It follows that 
\begin{align}
\vec s:=\bigoplus_{\boldsymbol{X}} \vec s_{\boldsymbol{X}}\ ,\qquad \vec s_{\boldsymbol{X}}:=(\vec Q_{\boldsymbol{X}})_{\mathcal{M}(\boldsymbol{X})}-(\vec Q_{\boldsymbol{Y}})_{\mathcal{M}(\boldsymbol{X})}\ 
\end{align}
is a Q-structure on $\mathcal M$.
The evaluation map 
\begin{align}
{\rm ev}:\mathcal{M}(\boldsymbol{X})\times \boldsymbol{X}\to \boldsymbol{Y}
\end{align}
generates a pull-back operation ${\rm ev}^\ast$ that
sends forms from $\boldsymbol{Y}$ to $\mathcal{M}(\boldsymbol{X})\times \boldsymbol{X}$, which composes with $\int_{\boldsymbol{X}}\mu_{\boldsymbol{X}}(\cdot)$ to a map $\int_{\boldsymbol{X}}\mu_{\boldsymbol{X}}{\rm ev}^\ast(\cdot)$ sending forms from $\boldsymbol{Y}$ to $\mathcal M_{\boldsymbol{X}}$.
We assume that 

\noindent i) each source is paired with a graded-symplectic structure $\omega_{\boldsymbol{Y}}$ on the target obeying
\begin{align}
{\rm deg}(\omega_{\boldsymbol{Y}})+{\rm deg}(\mu_{\boldsymbol{X}})=-1\ ,
\end{align}
such that $\int_{\boldsymbol{X}}\mu_{\boldsymbol{X}}{\rm ev}^\ast\omega_{\boldsymbol{Y}}$ contains a symplectic structure $\omega_{\mathcal{M}}$ on $\mathcal{M}$ of degree $-1$; and

\noindent ii) $\mathcal{M}$ has an S-structure inducing a $\Delta$ that annihilates\footnote{This can be achieved by introducing an auxiliary metric structure on the body of $\boldsymbol{X}$ facilitating i) a measure on the body $\mathcal{B}$ of $\mathcal{M}$ which induces a measure on $\mathcal{M}$ using Schwarz' fermionic Fourier transformation of Cartan integrals on submanifolds of $\mathcal{B}$ 
\cite{Schwarz:1992nx}; and ii) a point-splitting scheme on $\boldsymbol{X}$ that regularizes $\Delta$.} $\int_{\boldsymbol{X}}\mu_{\boldsymbol{X}}{\rm ev}^\ast(\cdot)$.

\noindent
Under these assumptions, 
\begin{align}
S|_{\mathcal{M}(\boldsymbol{X})}:=\int_{\boldsymbol{X}}\mu_{\boldsymbol{X}} \left(\imath_{\vec{Q}_{\boldsymbol{X}}}{\rm ev}^\ast\theta_{\boldsymbol{Y}}-{\rm ev}^\ast H_{\boldsymbol{Y}}\right)\ ,
\end{align}
where $\theta_{\boldsymbol{Y}}$ is the pre-symplectic one-form on the target, i.e., $\omega_{\boldsymbol{Y}}=d\theta_{\boldsymbol{Y}}$, defines a universal action $S$ on $\mathcal{M}$ that generates $\vec s$ in the bracket of $\omega_{\mathcal{M}}$, and is annihilated by $\Delta$.

\paragraph{Superconnections.}

The fundamental fields of the sigma model can be assembled into a Quillen superconnection 
\begin{align}
\boldsymbol{z}\in \boldsymbol{\mathcal{E}}_{[1]}\ ,\qquad \boldsymbol{\mathcal{E}}:=\Omega_{[0]}(\mathcal{M})\otimes \boldsymbol{\mathcal{V}}\ ,  
\end{align}
where $\boldsymbol{\mathcal {V}}$ is the $\mathbb{Z}$-graded vector space of homotopy associative algebra $(\boldsymbol{\mathcal{V}};\{m_r\}_{r=1}^\infty)$ with rank-$r$ products $m_r:\boldsymbol{\mathcal {V}}^{\otimes r}\to \boldsymbol{\mathcal {V}}$ of intrinsic degrees $2-r$. 
Letting $e_I\in \boldsymbol{\mathcal{V}}$ be basis elements of $\boldsymbol{\mathcal{V}}$, such that $\boldsymbol{v}\in \boldsymbol{\mathcal{E}}$ can be expanded as 
\begin{align}
\boldsymbol{v}=v^I \otimes e_I\ ,\qquad {\rm gh}(v^I)+{\rm deg}_{\boldsymbol{\mathcal{V}}}(e_I)={\rm totdeg}_{\boldsymbol{\mathcal E}}(\boldsymbol{v})\ ,
\end{align}
and $e^{\ast I}\in \boldsymbol{\mathcal{V}}^\ast$ be dual elements obeying $e^{\ast I}(e_J)=\delta^I_J$, hence ${\rm deg}_{\boldsymbol{\mathcal{V}}^\ast}(e^{\ast I})+{\rm deg}_{\boldsymbol{\mathcal{V}}}(e_I)=0$, one has 
\begin{align}
\boldsymbol{z}=z^I\otimes e_I\ ,\qquad z^I=e^{\ast I}(\boldsymbol{z})\ ,\qquad {\rm gh}(z^I)+{\rm deg}_{\boldsymbol{\mathcal{V}}}(e_I)=1\ ,
\end{align}
where $z^I$ is a set of graded-commutative coordinates on $\mathcal{M}$, i.e.,
\begin{align}
\Omega^{(\omega)}_{[0]}(\mathcal{M})\cong {\rm Env}((\boldsymbol{\mathcal{V}}[1])^\ast)\ ,
\end{align} 
the algebra of power series in $z^I$ with complex coefficients.

\paragraph{Noncommutative  geometries.} 
We take 
$\boldsymbol{\mathcal{V}}$ to be a Frobenius DGA represented by a space of functions\footnote{The space is defined up to similarity transformations generated by graded symmetric poly-vector fields \cite{Kontsevich:1997vb}.} on a graded manifold equipped with i) an associative composition rule; ii) a compatible Q-structure; and iii) a trace operation given by the Berezin integration measure.
Perturbative descriptions of such algebras arise by treating the functions as boundary functionals of first-quantized, two-dimensional AKSZ sigma models \cite{CaFe,Arias:2015wha} with targets containing differential Poisson manifolds $\boldsymbol{C}$ and additional gauge and internal sectors.
Non-perturbative completions, e.g., as algebras associated to metaplectic groups, yield
\begin{align}
\boldsymbol{\mathcal{V}}\cong \Omega_{[0]}(T[1]\boldsymbol{C})\otimes   \boldsymbol{\mathcal F}\ ,
\end{align}
where 
\begin{align} 
\boldsymbol{\mathcal F}\cong \mathbb{C}(\Gamma)\otimes \boldsymbol{\mathcal G}(K)\otimes \boldsymbol{\mathcal F}_{\rm int}
\end{align}
is a $\mathbb{Z}$-graded Frobenius DGA consisting of the algebra $\mathbb{C}(\Gamma)$ of a discrete group $\Gamma$ of differential Poisson maps $\gamma:\boldsymbol{C}\to \boldsymbol{C}$ encoding sectors into the operator algebra corresponding to imposing twisted boundary conditions on the two-dimensional sigma model; the algebra $\boldsymbol{\mathcal G}(K)$ generated by the ghost system of the continuous gauge group $K$ of the sigma model; and an algebra $\boldsymbol{\mathcal F}_{\rm int}$ generated by additional, internal sectors of the sigma model, c.f., the operator $\rho_{[-1]}$ of the four-dimensional BF-model.

\paragraph{Horizontal DGAs.} 

Assuming that the differential Poisson structure admits fermionic zero-modes induced by special fundamental vector fields generating fibers in $\boldsymbol{C}$ \cite{Arias:2016agc}, the resulting noncommutative   bundle\footnote{The special fundamental vector fields form the kernel of the push-forward of the projection map. } 
\begin{align}
\boldsymbol{F}\to \boldsymbol{C}\to \boldsymbol{M}
\end{align}
induces a horizontal DGA \cite{FSG1}
\begin{align}
\boldsymbol{\mathcal V}\cong \Omega_{[0]}(T[1]\boldsymbol{M})\otimes \boldsymbol{\mathcal{A}}\ ,\qquad \boldsymbol{\mathcal{A}}\equiv \Omega_{[0]}(\boldsymbol{F})\otimes \boldsymbol{\mathcal F}\ .
\end{align}
The higher products $m_r=0$, $r=3,4\dots$, and 
\begin{align}
m_1=\vec q_{T[1]\boldsymbol{M}}\otimes {\rm id}+{\rm id}\otimes \delta_{\boldsymbol{\mathcal{A}}}\ ,\qquad 
m_2=m_2^{(\boldsymbol{M})}\otimes m_2^{(\boldsymbol{\mathcal{A}})}\ ,
\end{align}
where $\vec q_{T[1]\boldsymbol{M}}$ is the canonical Q-structure on $T[1]\boldsymbol{M}$, i.e., $\boldsymbol{\mathcal V}$ is a parent module for various Cartan integrable subsystems arising via different consistent reductions achievable by either DGA-projections or homotopy contractions; in this context, $\boldsymbol{C}$ is referred as a correspondence space.

\paragraph{Equation of motion.}
The corresponding AKSZ data are
\begin{align}
\boldsymbol{X}{}&=T[1]\boldsymbol{M}\ ,\qquad \vec Q_{\boldsymbol{X}}=\vec q_{T[1]\boldsymbol{M}}\ ,\\\label{4.16}
\boldsymbol{Y}{}&=\boldsymbol{\mathcal{A}}[1]\ ,\qquad \vec Q_{\boldsymbol{Y}} \psi=\delta_{\boldsymbol{\mathcal{A}}}\psi+\psi^2\ ,\qquad \psi\in \boldsymbol{\mathcal{A}}[1]\ .
\end{align}
Thus, the superconnection
\begin{align}
\boldsymbol{z}|_{\boldsymbol{X}}\in \boldsymbol{\mathcal{E}}_{[1]}(\boldsymbol{X})\ ,\qquad \boldsymbol{\mathcal{E}}(\boldsymbol{X}):= \Omega_{[0]}(\mathcal{M}(\boldsymbol{X}))\otimes \Omega_{[0]}(\boldsymbol{X})\otimes \boldsymbol{\mathcal{A}}\ ,
\end{align}
and the classical equation of motion 
\begin{align}\label{2.26}
\boldsymbol{R}^{{\boldsymbol{z}}}:=\vec s_{\boldsymbol{X}} \boldsymbol{z}\approx 0\ ,
\end{align}
where the BV differential
\begin{align}
\vec s_{\boldsymbol{X}} \boldsymbol{z}= \vec q_{\boldsymbol{X}}\boldsymbol{z}- \delta_{\boldsymbol{\mathcal{A}}}\boldsymbol{z}-\boldsymbol{z}^2\ .
\end{align}

\paragraph{Fermion parities, degree maps, and Koszul convention.}  

Using the canonical DGA isomorphism 
\begin{align}
\Omega_{[0]}(\boldsymbol{X})\cong \Omega(\boldsymbol{M})\ ,\qquad \vec q_{\boldsymbol{X}}\cong d_{\boldsymbol{M}}\ ,
\end{align}
the DGA 
\begin{align}
\boldsymbol{\mathcal{E}}(\boldsymbol{X})\cong \boldsymbol{\mathcal{E}}(\boldsymbol{M})   
\end{align}
has six quantum numbers: 

\noindent i) second-quantized ghost number ${\rm gh}\equiv {\rm deg}_{\Omega_{[0]}(\mathcal{M})}\in\mathbb{Z}$;

\noindent ii) second-quantized fermion number ${\rm fer}_{\Omega_{[0]}(\mathcal{M})}\in \mathbb{Z}_2$;

\noindent iii) source degree ${\rm deg}_{\Omega_{[0]}(\boldsymbol{X})}={\rm deg}_{\Omega(\boldsymbol{M})}\in\{0,1,2,\dots\}$;

\noindent iv) source fermion parity ${\rm fer}_{\Omega_{[0]}(\boldsymbol{X})}={\rm fer}_{\Omega(\boldsymbol{M})}\in \mathbb{Z}_2$; 

\noindent v) first-quantized target degree ${\rm deg}_{{\mathcal A}}\in \mathbb{Z}$; and

\noindent vi) first-quantized target fermion parity ${\rm fer}_{{\mathcal A}}\in \mathbb{Z}_2$.

\noindent The superconnection $\boldsymbol{z}\in {\boldsymbol{\mathcal{E}}}(\boldsymbol{M})$ is assigned total degree, fermion parity and Grassmann bi-parity
\begin{align}
{\rm totfer}_{\boldsymbol{\mathcal{E}}(\boldsymbol{M})}(\boldsymbol{z})=0\ ,\qquad 
{\rm totdeg}_{\boldsymbol{\mathcal{E}}(\boldsymbol{M})}(\boldsymbol{z})=1 ,\qquad {\rm Gr}_{\boldsymbol{\mathcal{E}}(\boldsymbol{M})}(\boldsymbol{z})=(0,1)\ ,
\end{align}
defined as in \eqref{1.7}, \eqref{1.8}, and \eqref{1.9b}.
One also has 
\begin{align}
({\rm gh},{\rm fer}_{\Omega_{[0]}(\mathcal{M})},{\rm deg}_{\Omega(\boldsymbol{M})},{\rm fer}_{\Omega(\boldsymbol{M})},{\rm deg}_{{\mathcal A}},{\rm fer}_{{\mathcal A}})(d)=(0,0,1,0,0,0)\ ,\\ ({\rm gh},{\rm fer}_{\Omega_{[0]}(\mathcal{M})},{\rm deg}_{\Omega(\boldsymbol{M})},{\rm fer}_{\Omega(\boldsymbol{M})},{\rm deg}_{{\mathcal A}},{\rm fer}_{{\mathcal A}})(\vec s)=(1,0,0,0,0,0)\ ;
\end{align}
hence
\begin{align}
{\rm totfer}_{\boldsymbol{\mathcal{E}}}(d)=0\ ,\qquad 
{\rm totdeg}_{\boldsymbol{\mathcal{E}}}(d)=1 ,\qquad {\rm Gr}_{\boldsymbol{\mathcal{E}}}(d)=(0,1)\ ,\\
{\rm totfer}_{\boldsymbol{\mathcal{E}}}(\vec s)=0\ ,\qquad 
{\rm totdeg}_{\boldsymbol{\mathcal{E}}}(\vec s)=1 ,\qquad {\rm Gr}_{\boldsymbol{\mathcal{E}}}(\vec s)=(0,1)\ .
\end{align}
Thus, the Koszul sign convention \eqref{1.9} implies
\begin{align}
d\vec s+\vec sd=0\ ,
\end{align}
and that minus signs are produced when these differentials pass over odd elements of the superbundle, e.g., the superconnection.

\paragraph{Bases.}

Letting $\rho_a$ be a basis of $\boldsymbol{\mathcal{A}}$, and expanding
\begin{align}
\delta_{\boldsymbol{\mathcal{A}}}\rho_a=t_a{}^b \rho_b\ ,\qquad m_2(\rho_a,\rho_b)=t_{a,b}{}^c\rho_c\ ,
\end{align}
one has
\begin{align}
\psi=\psi^a \otimes \rho_a\ ,\qquad {\rm deg}(\psi^a)+{\rm deg}_{\boldsymbol{\mathcal{A}}}(\rho_a)=1\ ,
\end{align}
and $\vec Q_{\boldsymbol{Y}} \psi= (\vec Q_{\boldsymbol{Y}} \psi^a) \otimes \rho_a$ with
\begin{align}
\vec Q_{\boldsymbol{Y}} \psi^a= (-1)^{{\rm deg}(\psi^b)} \psi^b t_b{}^a+(-1)^{({\rm deg}(\psi^b)+1){\rm deg}(\psi^c)} \psi^b\psi^c m_{b,c}{}^a\ .   
\end{align}
Factorizing 
\begin{align}
e_I=\phi^r\otimes \rho_a\ ,\qquad \rho_a=\tau_i\otimes \sigma_\lambda\ ,
\end{align}
where $\phi^r\otimes \tau_i\otimes \sigma_\lambda$ is a basis for $\Omega_{[0]}(\boldsymbol{X})\otimes \Omega_{[0]}(\boldsymbol{F})\otimes \boldsymbol{\mathcal{F}}$, one may expand
\begin{align}
\boldsymbol{z}|_{\mathcal{M}(\boldsymbol{X})} = \boldsymbol{z}^\lambda \otimes \sigma_\lambda=Z^a \otimes \rho_a=z^a_r \otimes \phi^r\otimes \rho_a\ ,
\end{align}
where $\boldsymbol{z}^\lambda \in\Omega(\boldsymbol{M})\otimes \Omega_{[0]}(\boldsymbol{F})$ are horizontal forms on $\boldsymbol{C}$; $Z^a\in \Omega(\boldsymbol{M})$ are component forms on $\boldsymbol{M}$; and $z^a_r$ are graded commutative coordinates of $\mathcal{M}(\boldsymbol{M})$ providing the expansion coefficients of $z^a$ in the basis of forms $\phi^r$ on $\boldsymbol{M}$.

\paragraph{Homotopy contraction.} The equations of motion \eqref{2.26} admit solution spaces subject to boundary conditions obtainable by fixing i) the group of similarity transformations of $\boldsymbol{\mathcal{V}}$ by choosing a basis for the space of symbols of $\boldsymbol{\mathcal{V}}$; ii) a fibration $\boldsymbol{X}\stackrel{\hookleftarrow}{\rightarrow}\check{\boldsymbol{X}}$ over a graded commutative supermanifold $\check{\boldsymbol{X}}\equiv T[1]\check{\boldsymbol{M}}$; and iii) a fibre DGA, i.e., a set of symbols on the fibre space of the fibration closed under the associative composition rule and $\vec q_{\boldsymbol{X}}$.
These data induce a perturbatively defined
homotopy contraction to a (universally) Cartan integrable reduced system 
\begin{align}\label{2.35}
\check{\boldsymbol{R}}^{\check{\boldsymbol{z}}}:= d_{\check{\boldsymbol{M}}}\check{\boldsymbol{z}}- \sum_{r=1}^\infty  \check{l}_r(\check{\boldsymbol{z}}^{\wedge r})\approx 0\ ,\qquad \check{\boldsymbol{z}}\in \check{\boldsymbol{\mathcal V}}\cong \Omega(\check{\boldsymbol{M}})\otimes \check{\boldsymbol{\mathcal{L}}}\ , 
\end{align}
where $(\check{\boldsymbol{\mathcal L}},\{\check{l}_{r=1}^\infty\})$ is a homotopy Lie algebra with vector space\footnote{Thus, $\check{\boldsymbol{\mathcal L}}$ and $\boldsymbol{\mathcal A}$ share the same basis, as in \eqref{C.16}.}
\begin{align}\label{2.36}
\check{\boldsymbol{\mathcal L}}\stackrel{\rm vec}{\cong}\boldsymbol{\mathcal A}\ ,
\end{align}
and graded brackets\footnote{The homotopy Lie algebra $(\check{\boldsymbol{\mathcal L}},\{\check{l}_{r=1}^\infty\})$ admits an extension to a homotopy associative algebra $(\check{\boldsymbol{\mathcal A}},\{\check{m}_{r=1}^\infty\})$ with vector space $\check{\boldsymbol{\mathcal A}}\cong \check{\boldsymbol{\mathcal L}}$ and homotopy associative graded products $\check{m}_r$, $r=1,2,\dots$, of degrees $2-r$ whose graded anti-symmetrizations define $\check{l}_r$ \cite{Sharapov:2019vyd}.} $\check{l}_r$, $r=1,2,\dots$, of degrees $2-r$ obeying homotopy Jacobi identities implying Bianchi identities
\begin{align}
d_{\check{\boldsymbol{M}}}\check{\boldsymbol{R}}^{\check{\boldsymbol{z}}}+\sum_{r=1}^\infty 
 r \check{l}_r(\check{\boldsymbol{z}}^{\wedge (r-1)}\wedge \check{\boldsymbol{R}}^{\check{\boldsymbol{z}}})\equiv 0\ .
\end{align}
It follows that the Cartan curvatures transform homogeneously under the gauge transformations 
\begin{align}
\delta_{\boldsymbol{\epsilon}}\check{\boldsymbol{z}}=d_{\check{\boldsymbol{M}}}\check{\boldsymbol{\epsilon}}+\sum_{r=1}^\infty r\check{l}_r(\check{\boldsymbol{z}}^{\wedge (r-1)}\wedge \check{\boldsymbol{\epsilon}})\ ,
\end{align}
with unconstrained gauge parameters $\check{\boldsymbol{\epsilon}}$, viz.,
\begin{align}
\delta_{\boldsymbol{\epsilon}}\check{\boldsymbol{R}}^{\check{\boldsymbol{z}}}=\sum_{r=1}^\infty 
 r (r-1) \check{l}_r(\check{\boldsymbol{z}}^{\wedge (r-2)}\wedge \check{\boldsymbol{\epsilon}}\wedge \check{\boldsymbol{R}}^{\check{\boldsymbol{z}}})\ ;
\end{align}
finite gauge transformations yield classical moduli spaces with the structure of Lie groupoids, referred to as  Cartan integration modules of the contracted system.

\paragraph{Lie-algebra modules.} 

Denoting the $\check\ell_1$-cohomology\footnote{Passing to $\check{\boldsymbol{\mathcal H}}$ amounts to contracting the Stuckelberg pairs from $\boldsymbol{z}$.} in $\check{\boldsymbol{{\mathcal L}}}$ by $\check{\boldsymbol{{\mathcal H}}}$, and decomposing $\check{\boldsymbol{{\mathcal H}}}=\bigoplus_{n\in \mathbb{Z}}\check{\boldsymbol{{\mathcal H}}} _{[n]}$ under degree, it follows that
\begin{align}
\mathfrak{g}:=(\check{\boldsymbol{{\mathcal H}}}_{[0]},\check{l}_2)
\end{align}
is a graded Lie algebra represented in $\check{\boldsymbol{{\mathcal H}}}_{[n]}$ by\footnote{In the semi-classical theory, $x\in \mathfrak{g}$ lifts to a vector fields $\vec x$ on ${\mathcal M}$ defined by $\vec x(\boldsymbol{z})+\check{l}_2(x,\boldsymbol{z}):=0$.}
\begin{align} \check l_2:\mathfrak{g}\otimes \check{\boldsymbol{{\mathcal H}}}_{[n]}\to \check{\boldsymbol{{\mathcal H}}}_{[n]}\ .\end{align}
In the semi-classical theory, $x\in \mathfrak{g}$ lifts to a vector fields $\vec x$ on ${\mathcal M}$ defined by 
\begin{align}
\vec x(\boldsymbol{z}):= -\check{l}_2(x,\boldsymbol{z})\ ,
\end{align}
which we use to assign component fields conformal weights; see Section \ref{Subsection3formcocycle} and \eqref{A.26}.

\section{Superconformal Algebraic Structures}\label{Sec:2}
\label{algebraicstructures}

This section describes Lie algebras, groups, and modules arising in the unfolded $(2,0)$ model. 
Inspired by properties of recently proposed conformal duals of Vasiliev-like higher-spin gravities \cite{FSG1}, the basic assumption is that the model admits 

i) a non-perturbative formulation in terms of conformal group modules; and 

ii) perturbative expansions around classical backgrounds in terms of Lie algebra modules consisting of tensors of an unbroken structure group.

\noindent 

The formalism provides an algebraically satisfactory treatment of classical singularities, whereby the group modules of (i) contain unitarizable submodules mapped by monomorphisms into subspaces of the Lie algebra modules of (ii) consisting of regular configurations, e.g., plane waves in locally conformal Minkowski spacetime and bounded modes on $S^1\times S^5$.
The group modules of (i) can be extended to include non-normalizable states mapped to configurations in (ii) that are singular on submanifolds of Cartan integration manifolds, e.g., unfolded boundary-to-bulk propagators, localized at the boundary, and unfolded Coulomb-like singularities localized inside the bulk \cite{BTZ2019,corfu19,FSG1}.

The local degrees of freedom of unfolded systems are encoded into infinite-dimensional zero-form modules.
More precisely, (ii) yields zero-form modules containing primary curvature and matter tensors obeying Bargmann-Wigner equations.
The formalism naturally includes dual zero-form modules \cite{BMVI} containing dual curvature tensors obeying dual Bargmann-Wigner equations.
At the linearized level, these modules share unitarizable degrees of freedom on-shell;
in conformal contexts \cite{FSG1}, they are related by Fourier transformations in twistor spaces.
In the context of the (2,0) model, its zero-form module contains the chiral curvature tensor, and its dual contains a dual field strength with the opposite chirality obeying a dual equation of motion, such that vertices including both zero-forms may produce spacetime non-localities as expected in the $(2,0)$ model \cite{DouglasTalk}.

\subsection{Superalgebra}

\paragraph{Conformal decomposition.}

The unfolded system's symmetry algebra $\mathfrak{g}$ contains a dilation operator $D$.
Assuming that $D$ can be diagonalized in the unfolding module using purely imaginary eigenvalues, the corresponding basis elements obey
\begin{align}\label{3.1} \rho(D,\tau_{i})=i\Delta_i \tau_{i}\ ,\qquad \Delta_i\in \mathbb{R}\ ,
\end{align}
which we denote by $\tau_i\equiv \tau_{\Delta_i}$.
The symmetry algebra is the superconformal algebra
\begin{align}
\mathfrak{g}\equiv \mathfrak{osp}(8^\ast|4)=\mathfrak{so}(2,6)_{\rm Conf}\oplus \mathfrak{usp}(4)_{\rm R}\oplus \mathfrak{q}^{(+)}_{+1/2}\oplus \mathfrak{q}^{(-)}_{-1/2}\ , 
\end{align}
where $\mathfrak{q}^{(\pm)}_{\pm1/2}$ are fermionic subspaces spanned by chiral supercharges $Q^{(\pm)}_{\pm 1/2}$.
The conformal subalgebra 
\begin{align}
\mathfrak{so}(2,6)_{\rm Conf}\downarrow_{\mathfrak{so}(1,5)_{\rm Lor}}= \mathfrak{so}(1,5)_{\rm Lor}\oplus \mathfrak{t}_{+1}\oplus \mathfrak{t}_{-1}\oplus \mathfrak{so}(1,1)_{\rm Dil}\ ,
\end{align}
where $\mathfrak{t}_{\pm 1}$ are spanned by $\mathfrak{so}(1,5)_{\rm Lor}$-vectorial translation operators $T_{\pm 1}$ whose commutators close on $\mathfrak{so}(1,5)_{\rm Lor}\oplus  \mathfrak{so}(1,1)_{\rm Dil}$. 
Under $\mathfrak{so}(1,1)_{\rm Dil}$, the superconformal algebra acquires a five-grading, viz.,
\begin{align}
\mathfrak{g}=\bigoplus_{\Delta\in \{-1,-1/2,0,1/2,1\}} \mathfrak{g}_{\Delta}\ ,
\end{align}
where $\mathfrak{g}_{\pm 1}\equiv \mathfrak{t}_{\pm1}$,  $\mathfrak{g}_{\pm 1/2}\equiv \mathfrak{q}^{(\pm)}_{\pm1/2}$, and 
\begin{align}
\mathfrak{g}_{0}\equiv \mathfrak{h}:=\mathfrak{so}(1,5)_{\rm Lor}\oplus \mathfrak{so}(1,1)_{\rm Dil}\oplus \mathfrak{usp}(4)_{\rm R}\ ,
\end{align}
to be treated as the Lie algebra of the structure group of the model.
The subalgebras 
\begin{align}
{\mathfrak{g}}_{\pm}:= \bigoplus_{\Delta\in \{0,\pm 1/2,\pm 1\}} \mathfrak{g}_{\Delta}
\end{align}
are two distinct, isomorphic, copies
of the semi-direct sum of $\mathfrak{so}(1,1)_{\rm Dil}$ and the $(2,0)$ super-Poincar\'e algebras with chiral supercharges $Q^{(\pm)}_{\pm 1/2}$, i.e.\footnote{We let $\mathfrak{a}\oplus_\sigma \mathfrak{b}$ stand for the semi-direct sum of two Lie algebras $\mathfrak{a}$ and $\mathfrak{b}$ with $[a,b]=\rho(a)b$ where $\rho$ is a representation of $\mathfrak{a}$ in $\mathfrak{b}$.},
\begin{align}
{\mathfrak{g}}_{+}\equiv \mathfrak{so}(1,1)_{\rm Dil}\oplus_{\Delta} \mathfrak{iso}(1,5|16,0)\ ,\qquad {\mathfrak{g}}_{-}\equiv \mathfrak{so}(1,1)_{\rm Dil}\oplus_{\Delta} \mathfrak{iso}(1,5|0,16)\ .
\end{align}

\paragraph{Graded commutation rules in $\mathfrak{h}$-covariant basis.} We let  $(M_\alpha{}^\beta,D,N_{IJ})$ denote the generators of $\mathfrak{h}$ with commutation rules
\begin{align}
&[M_{\alpha}{}^{\beta},M_{\gamma}{}^{\delta}]=i(\delta_\gamma^\beta M_{\alpha}{}^{\delta}-\delta_\alpha^\delta M_{\gamma}{}^{\beta}), &[N_{IJ},N_{KL}]={}4i\eta_{(J|(K}N_{L)|I)}\ ,\\
&[D,M_{\alpha}{}^{\beta}]=0\ , &[D,N_{IJ}]=0\ .
\end{align}
In a ${\mathfrak{g}}_{+}$-background, the generators of translations, special conformal transformations, supersymmetries, and superconformal transformations are denoted by
\begin{align}
T_{\alpha\beta}\equiv T_{\alpha\beta; +1}\ ,\qquad K^{\alpha\beta}\equiv T^{\alpha\beta}_{-1}\ ,\qquad 
Q_{\alpha I}\equiv Q^{(+)}_{\alpha I;+1/2}\ ,\qquad S^{\alpha }{}_I\equiv Q^{(-)\alpha }{}_{I;-1/2}\ ,
\end{align}
respectively, where thus $T_{\alpha\beta}, Q_{\alpha I}\in {\mathfrak{g}}_{+}$ and $K^{\alpha\beta},S^{\alpha}{}_I\in \mathfrak{g}/{\mathfrak{g}}_{+}$.
The Lorentz spins and conformal weights of the translation and special conformal generators are encoded into
\begin{align}
&[M_{\alpha}{}^{\beta},T_{\gamma\delta}]={}2i(\delta_{[\gamma|}^\beta T^{\phantom{\beta}}_{\alpha|\delta]}-\frac14 \delta_\alpha^\beta T_{\gamma\delta})\ ,\qquad 
[M_{\alpha}{}^{\beta},K^{\gamma\delta}]=-2i(\delta_{\alpha}^{[\gamma|} K^{\beta|\delta]}_{\phantom{|}}-\frac14 \delta_\alpha^\beta K^{\gamma\delta})\ ,\\
&[D,T_{\alpha\beta}]=iT_{\alpha\beta}\ ,\qquad [D,K^{\alpha\beta}]=-iK^{\alpha\beta}\ .
\end{align}
The conformal algebra is closed by taking 
\begin{align}
[T_{\alpha\beta},K^{\gamma\delta}]={}&4i\delta_{[\beta}^{[\gamma}(M_{\alpha]}^{\delta]}+\frac12\delta_{\alpha]}^{\delta]} D)\ .\end{align}
Taking the Lorentz tensorial basis of the conformal algebra to be
\begin{align}
T_a:= \frac12 (\tilde\sigma_a)^{\alpha\beta}T_{\alpha\beta}\ ,\qquad M_{ab}:=-\frac12 (\sigma_{ab})_\alpha{}^\beta M_\beta{}^\alpha\ ,\qquad K_a:=-\frac12 (\sigma_a)_{\alpha\beta}K^{\alpha\beta}\ ,
\end{align}
yields the ``yellow-book'' conventions
\begin{align}
&[M_{ab},M_{cd}]={}4i\eta_{[c|[b]}M_{a]|d]}\ ,\qquad [M_{ab},T_{c}]=2i\eta_{c[b}T_{a]}\ ,\qquad [M_{ab},K_{c}]=2i\eta_{c[b}K_{a]}\ ,\\
&[D,T_a]={}iT_a\ ,\qquad [D,K_a]=-iK_a\ ,\\
&[K_a,T_b]=2i(\eta_{ab}D-M_{ab})\ ;
\end{align}
conversely,
\begin{align}
T_{\alpha\beta}:= -\frac12(\sigma^a)_{\alpha\beta} T_a\ ,\qquad M_\alpha{}^\beta=\frac14 (\sigma^{ab})_\alpha{}^\beta M_{ab}\ ,\qquad K^{\alpha\beta}:= \frac12(\tilde{\sigma}^a)^{\alpha\beta} K_a\ .
\end{align}
The transformation properties of the supercharges under the structure group are encoded into 
\begin{align}
[M_\alpha{}^\beta,Q_{\gamma I}]={}&i(\delta_\gamma^\beta Q_{\alpha I}-\frac14\delta_\alpha^\beta Q_{\gamma I})\ ,\qquad [M_\alpha{}^\beta,S^{\gamma}{}_I]=-i(\delta^\gamma_\alpha S^{\beta}{}_I-\frac14\delta_\alpha^\beta S^{\gamma}{}_I)\ ,\\
[D,Q_{\alpha I}]={}&\frac{i}2 Q_{\alpha I}\ ,\qquad [D,S^{\alpha}{}_I]=-\frac{i}2 S^{\alpha}{}_{I}\ ,\\
[N_{IJ}, Q_{\alpha K}]={}&2i \eta_{(I|K} Q_{|\alpha J)}\ ,\qquad
[N_{IJ},S^{\alpha}{}_K]=2i \eta_{(I|K} S^{\alpha}{}_{|J)}
\ .
\end{align}
The superconformal algebra is closed by 
the anti-commutators\footnote{In the $(2,2)$ super-Poincar\'e algebra, one instead has $[Q,Q]=T=[S,S]$ and $[Q,S]=0$.}
\begin{align}\label{qqssanticommutators}
[Q_{\alpha I},Q_{\beta J}]={}&\eta_{IJ} T_{\alpha\beta}\ ,\qquad [S^{\alpha}{}_I,S^{\beta}{}_{J}]=\eta_{IJ} K^{\alpha\beta} \ ,\\
\label{QSanticommutator}
[Q_{\alpha I},S^{\beta}{}_J]={}&\eta_{IJ} \left(M_{\alpha}{}^{\beta}+\frac12\delta_\alpha^\beta D\right)+\delta_\alpha^\beta N_{IJ}\ ,
\end{align}
and commutators 
\begin{align}
[T_{\alpha\beta},S^\gamma{}_I]={}&2i\delta_{[\alpha}^\gamma Q^{\phantom{\gamma}}_{\beta]I}\ ,\qquad [K^{\alpha\beta},Q_{\gamma I}]=-2i\delta_{\gamma\phantom{I}}^{[\alpha} S^{\beta]}{}_I\ ,
\end{align}
while $[T,Q]=0=[K,S]$.

\subsection{Supersingleton} \label{Sec:supersingleton}

\paragraph{Algebraic equation of motion.} 

The enveloping algebra $\boldsymbol{{\rm Env}}(\mathfrak{g})$, i.e., the space of polynomials in the generators of $\mathfrak{g}$ modulo the graded commutation rules including a unital element, contains a two-sided ideal 
\begin{align}
\boldsymbol{\mathcal{I}}(V):=V\boldsymbol{{\rm Env}}(\mathfrak{g})=\boldsymbol{{\rm Env}}(\mathfrak{g}) V\ ,   
\end{align}
generated by a subspace $V$ spanned by a set of quadratic elements (including the quadratic Casimir), inducing a unital, associative algebra 
\begin{align}
\boldsymbol{\mathcal{HS}}(V):={\rm Env}(\mathfrak{g})/\boldsymbol{\mathcal{I}}(V)\ ,
\end{align}
decomposing under ${\rm ad}_{\mathfrak{g}}$ into an infinite tower of finite-dimensional $\mathfrak{g}$-multiplets consisting of states Lorentz spins in finite interval.
The algebra $\boldsymbol{\mathcal{HS}}(V)
$ admits dual realizations as as: a) the gauge algebra of seven-dimensional ${\cal N}=2$ higher-spin supergravity \cite{Sezgin:2002rt}; b) the gauge algebra of a yet-to-be constructed six-dimensional, topological ${\cal N}=(2,0)$ conformal higher-spin supergravity coupled to conformal matter analogously to \cite{FSG1,FSG2}; and c) as the algebra of functions on the phase space of the conformal superparticle with algebraic equation of motion $V\approx 0$, arising as the supermembrane parton in $AdS_7\times S^4$ \cite{Engquist:2005yt}. 

\paragraph{Algebra and group modules.}
The algebra $\boldsymbol{\mathcal{HS}}(V)
$ admits two irreducible representations in a pair of Hermitian spaces $\mathsf{T}^{(\pm)}$ that i) are dual as $\mathfrak{g}$-modules, viz., 
\begin{align}
\mathsf{T}^{(\pm)}\cong (\mathsf{T}^{(\mp)})^\ast\ ;
\end{align}
and ii) decompose under $\mathfrak{h}$ into finite-dimensional $\mathfrak{h}$-irreps, using conventions\footnote{In our conventions, the basis elements of $T^{(+)}$ have lower spinor indices.} which yield tensors and spinors that are selfdual and left-handed, respectively, in the case of (+), and anti-selfdual and right-handed in the case of (-).
Thus, $\mathsf{T}^{(\pm)}$ are acted on faithfully by the structure group
\begin{align}
G_{\mathfrak{h}}:=SO(1,5)_{\rm Lor}\times SO(1,1)_{\rm Dil}\times USp(4)_{\rm R}\ .
\end{align} 
Starting from the conformal supergroup 
\begin{align}
G\equiv OSp(8^\ast|4)\ ,
\end{align}
it is possible to characterize the \emph{supersingleton} $\mathsf{S}$ as the minimal, hermitian, selfdual $\boldsymbol{\mathcal{HS}}(V)$-module that is also a projective $G$-module, i.e., $G$ acts faithfully on $\mathsf{S}$ in a projective representation, and 
\begin{align}
\mathsf{S}\cong \mathsf{S}^\ast\ ,
\end{align}
in the sense that if $\mathsf{R}$ is a projective $G$-module and $\mathsf{R}\subseteq \mathsf{S}$, then $\mathsf{R}^\ast \subseteq \mathsf{S}$.
The supersingleton contains a dual pair of $\mathfrak{g}$-submodules 
\begin{align}
\mathsf{S}^{(\pm)}=(\mathsf{S}^{(\mp)})^\ast\subset \mathsf{S}\ , 
\end{align}
equipped with monomorphisms\footnote{Loosely speaking, $\mathsf{S}^{(\pm)}$ consist of non-polynomial elements while $\mathsf{T}^{(\pm)}$ consist of formal power series in the generators of $\mathfrak{g}$. Indeed, $\mathsf{S}^{(\pm)}$ can be built via the one-sided action of $\boldsymbol{{\rm Env}}(\mathfrak{g})$ on a reference element belonging to $Mp_{\infty}(G^{\Comp})$, the asymptotic boundary of $Mp(G^{\Comp})$, which is a two-sided module of the complex metaplectic group $Mp(G^{\Comp})$ containing states with distinct left- and right-polarizations \cite{meta,FSG1}; see Conclusions, for further comments.} 
\begin{align}\label{3.32}
\tau^{(\pm)}:\mathsf{S}^{(\pm)}\to \mathsf{T}^{(\pm)}\ ,
\end{align}
rendering $\mathsf{S}$ meanings as non-polynomial completions of $\mathsf{T}^{(+)}\oplus \mathsf{T}^{(-)}$  \cite{fibre}. 

\subsection{Harmonic analysis}

The unfolded formalism provides an algebraic approach to harmonic analysis \cite{fibre}, i.e., constructing spaces of linearized, classical solutions on super-Minkowski backgrounds subject to various conditions forming projective $G$-modules $\mathsf{S}_\infty(\xi)$, referred to as extended singletons, labelled by coset elements $\xi\in \mathfrak{g}/{\rm Ad}_G$, corresponding, in the different realizations outlined in Section \ref{Sec:supersingleton}, to boundary conditions on free fields of (a) and (b), and polarizations of the conformal superparticle of (c).

\paragraph{Boundary conditions.}

On super-Minkowski spacetimes $(\boldsymbol{M}_{(6|16)},{\boldsymbol{\Omega}}_{\pm})$, i.e, manifolds $\boldsymbol{M}_{(6|16)}$ of superdimension $(6|16)$ equipped with ${\mathfrak{g}}_{\pm}$-valued one-forms ${\boldsymbol{\Omega}}_{\pm}$ obeying 
\begin{align}
d\boldsymbol{\Omega}_{\pm}+\boldsymbol{\Omega}_{\pm}\wedge \boldsymbol{\Omega}_{\pm}\approx 0\ ,
\end{align}
the unfolding of an abelian two-form superpotential with selfdual (+) or anti-selfdual (-) three-form field strength on the body of $\boldsymbol{M}_{(6|16)}$  yield chiral zero-forms $\boldsymbol{h}^{(\mp)}$ valued in $\mathsf{T}^{(\mp)}\downarrow_{\mathfrak{g}_\pm}$, and dual chiral zero-forms $\boldsymbol{h}^{(\pm)}$ valued in  $\mathsf{T}^{(\pm)}\downarrow_{\mathfrak{g}_\pm}$, descending from an $\mathsf{S}$-valued parent zero-form $\boldsymbol{h}$, i.e., 
\begin{align}
\boldsymbol{h}^{(\pm)}=\tau^{(\pm)}(\boldsymbol{h}|_{\mathsf{S}^{(\pm)}})\ ,
\end{align}
which is globally, covariantly constant, i.e.,
\begin{align}
d\boldsymbol{h} +\boldsymbol{\Omega}_{\pm}\boldsymbol{h}\approx 0\ ,
\end{align}
without any source\footnote{The $\tau^{(\pm)}$-projected equations of motion for $\boldsymbol{h}^{(\pm)}$ may contain sources; for a related discussion, see \cite{BTZ2019}.}.
Viewed as a hermitian vector space, the supersingleton $\mathsf{S}$ admits extensions by points at its infinity forming $G$-irreps $\mathsf{S}_\infty(\xi)$ labelled by superalgebra coset elements $\xi$ corresponding to distinct boundary conditions\footnote{
In the geometric quantization of symplectic manifolds, irreducible Hilbert spaces arise by choosing polarizations.
While $\mathbb{R}^{2n}$ topology only leaves one possibility (up to unitary equivalences), nontrivial topologies allow multitudes of choices.
It has been proposed \cite{FSG1} that noncommutative   geometries described by Wigner-deformed oscillators (see, e.g., \cite{Vasiliev:1989re}) arise as solutions to Vasiliev-type higher-spin gravities obtained by perturbative completions of linearized configurations from $S_\infty$ corresponding holographically to conformal theories in different topologies \cite{BTZ2019,FSG1}.} \cite{meta,FSG1,FSG2}.
For example, incoming and outgoing waves on conformal Minkowski spacetimes arise in irreducible, massless, Wigner mass-shells $\mathsf{W}^\pm\subset \mathsf{S}$ with strictly positive (+) and negative (-) Poincaré energies.
The restrictions of the Hermitian form of $\mathsf{S}$ to $\mathsf{W}^\pm$ are definite, i.e., $\mathsf{W}^\pm$ are unitarizable hence module selfdual, viz.,
\begin{align}
\mathsf{W}^\pm\cong (\mathsf{W}^\pm)^\ast\ .
\end{align}
Likewise, waves on conformal $S^1\times S^5$ spacetime expanded over spherical harmonics on $S^5$ yield massless irreps $\mathsf{D}^\pm\subset \mathsf{S}$ inside compact weight spaces with strictly positive (+) and negative (-) AdS energies, such that 
\begin{align}
\mathsf{D}^\pm\cong (\mathsf{D}^\pm)^\ast\ ,  
\end{align}
and $\mathsf{D}^\pm$ are separately
unitarizable.

\paragraph{Automorphisms.}

The superconformal algebra and group admit several outer automorphisms\footnote{The coset construction of superspace extension of $AdS_7\times S^4$ makes use of an automorphism group given by ${\mathbb Z}_4$ that stabilizes $\mathfrak{so}(1,6)\oplus 
\mathfrak{so}(4)$. 
The invariant $\mathfrak{so}(1,6)$ is generated by
$\{M_{ab}, \frac12(T_{\alpha\beta}-K_{\alpha\beta})\}$, and the invariant $\mathfrak{so}(4)\cong \mathfrak{su}(2)\oplus\mathfrak{su}(2)$ appears in a decomposition $\mathfrak{usp}(4)=
\mathfrak{su}(2)\oplus \mathfrak{su}(2) \oplus(\mathfrak{su}(2)\oplus\mathfrak{su}(2))^\perp$. 
} .
The superalgebra automorphisms\footnote{These two automorphisms were introduced in \cite{FSG1} using a different notation adapted to an oscillator realization: $\pi_\s$ was denoted by $\pi_{\!\mathscr{F}}$, and $\pi_\kappa$ by $\pi_{\!\mathscr{P}}$.} 
\begin{align}
\pi_\sigma:\mathfrak{g}\to\mathfrak{g}\ ,\qquad (\pi_\sigma)^2={\rm Id}_{\mathfrak{g}}\ ,\\
\pi_\kappa:\mathfrak{g}\to\mathfrak{g}\ ,\qquad (\pi_\kappa)^4={\rm Id}_{\mathfrak{g}}\ ,    
\end{align}
extend to automorphisms of the superconformal group given by\footnote{The $\mathfrak{g}$-automorphisms induce maps from the image ${\rm Exp}(\mathfrak{g})\subset G$ to itself, which extend to $G$-automorphisms.} 
\begin{align}
\Pi_\sigma:G\to G\ ,\qquad (\Pi_\sigma)^2={\rm Id}_G\ ,\\
\Pi_\kappa:G\to G\ ,\qquad (\Pi_\kappa)^4={\rm Id}_G\ ,
\end{align}
and corresponding endomorphisms
\begin{align}
\sigma:\mathsf{S}\to\mathsf{S}\ ,\qquad \sigma^2=1\ ,\\
\kappa:\mathsf{S}\to\mathsf{S}\ ,\qquad \kappa^4=1\ ,
\end{align}
such that
\begin{align}
\sigma x \psi=\pi_\sigma(x)\sigma\psi\ ,\qquad \kappa x \psi=\pi_\kappa(x)\kappa\psi\ ,
\end{align}
for $x\in \boldsymbol{\mathcal{A}}$ and $\psi\in \mathsf{S}$, and 
\begin{align}
{}&\sigma:\mathsf{S}^{(\pm)}\to \mathsf{S}^{(\mp)}\ ,\\
{}&\kappa: \mathsf{W}^{\pm}\to \mathsf{W}^{\mp}\ ,\quad  \mathsf{D}^{\pm}\to \mathsf{D}^{\mp}\ .
\end{align}
The $\mathfrak{g}$-automorphism 
\begin{align}
{}&\pi_\sigma:\mathfrak{t}_{\pm 1}\to  \mathfrak{t}_{\mp 1}\ ,\qquad \pi_\sigma:\mathfrak{q}_{\pm 1/2}\to  \mathfrak{q}_{\mp 1/2}\ ,\\
{}&\pi_\sigma|_{\mathfrak{h}}={\rm Id}_{\mathfrak{so}(1,5)_{\rm Lor}}\oplus (-{\rm Id}_{\mathfrak{so}(1,1)_{\rm Dil}})\oplus{\rm Id}_{\mathfrak{usp}(4)_{\rm R}} \ ,
\end{align}
is given in the $\mathfrak{h}$-covariant basis by\footnote{Defining new gamma matrices $(\sigma^{\prime a})_{\alpha\beta}=(\tilde\sigma^{ a})^{\alpha\beta}$ and $(\tilde\sigma^{\prime a})^{\alpha\beta}=(\sigma^{ a})_{\alpha\beta}$, the new chirality matrix $(\Gamma')_{\underline\alpha}{}^{\underline\beta}=-(\Gamma)_{\underline\alpha}{}^{\underline\beta}$, such that $\pi_\sigma(\Gamma Q_I)=\Gamma' \pi_\sigma(Q_I)$ and $\pi_\sigma(\Gamma S_I)= \Gamma'\pi_\sigma(S_I)$.}
\begin{align}
\pi_\sigma(T_{\alpha\beta})=K^{\alpha\beta}\ ,\qquad \pi_\sigma(K^{\alpha\beta})=T_{\alpha\beta}\ ,\qquad 
\pi_\sigma(Q_{\alpha I})=S^{\alpha}{}_{I}\ ,\\
\pi_\sigma(M_\alpha{}^\beta)=-M_\beta{}^\alpha\ ,\qquad \pi_\sigma(D)=-D\ ,\qquad \pi_\sigma(N_{IJ})=N_{IJ}\ ;
\end{align}
it is broken spontaneously by super-Poincaré backgrounds.
The $\mathfrak{g}$-automorphism 
\begin{align}
\pi_\kappa:\mathfrak{t}_{\pm 1}\to  \mathfrak{t}_{\pm 1}\ ,\qquad \pi_\kappa:\mathfrak{q}_{\pm 1/2}\to  \mathfrak{q}_{\pm 1/2}\ ,\qquad
\pi_\kappa|_{\mathfrak{h}}={\rm Id}_{\mathfrak{h}} \ ,
\end{align}
is given in the $\mathfrak{h}$-covariant basis by 
\begin{align}
\pi_\kappa(T_{\alpha\beta}){}&=-T_{\alpha\beta}\ ,\qquad \pi_\kappa(K^{\alpha\beta})=-K^{\alpha\beta}\ ,\\
\pi_\kappa(Q_{\alpha I}){}&=iQ_{\alpha I}\ ,\qquad \pi_\kappa(S^{\alpha}{}_{I})=-iS^{\alpha}{}_{I}\ ;
\end{align}
it is left unbroken by super-Poincar\'e backgrounds and its restriction to $\mathfrak{so}(2,6)$ is involutive.

\section{Vacuum Connections}
\label{background}

\begin{table}[t]
\centering
\begin{tabular}{|c|c|c|c|c|}
\hline
Component & ${\rm totdeg}_{\boldsymbol{\mathcal{E}}(\boldsymbol{M})}$  & ${\rm totfer}_{\boldsymbol{\mathcal{E}}(\boldsymbol{M})}$ & ${\rm Gr}_{\boldsymbol{\mathcal{E}}(\boldsymbol{M})}$ & $\Delta_{\Omega_{[0]}(\mathcal{M})}$ \\
field & $\in \mathbb{Z}$ & $\in \mathbb{Z}_2$ & $\in \mathbb{Z}_2\times \mathbb{Z}_2$ &$\in \mathbb{R}$ \\
\hline
$E^{\alpha\beta}$ & $1$ & $0$ & $(1,0)$ & $-1$\\
$F^{\alpha I}$ & $1$ & $1$ & $(1,1)$ &$-1/2$\\
$\widetilde{E}_{\alpha\beta}$ & $1$ & $0$ & $(1,0)$ & $+1$\\
$\widetilde{F}_{\alpha}{}^I$ & $1$ & $1$ & $(1,1)$ & $+1/2$\\
\hline
\end{tabular}
\caption{Quantum numbers of background frame fields.}
\label{tab:1}
\end{table}

This section describes flat superconnections on (super)manifolds of arbitrary (super)dimensions, which enter the construction in Sections 5 and 6 of the universal Cartan integrable system yielding superspace as well as component formulations depending on choices of background \cite{Engquist:2002gy,Grigoriev:2025vsl}; see also Appendix \ref{App:symmetriescocycles}.

\subsection{Superconformal background}

Working on a source $\boldsymbol{X}=T[1]\boldsymbol{M}$ where $\boldsymbol{M}$ is a (super)manifold, the superconformal background connection is an element 
\begin{align}
{\boldsymbol{\Omega}}\in \Omega_{[0]}(\mathcal{M})\otimes \Omega(\boldsymbol{M})\otimes \mathfrak{g}\ ,
\end{align} 
with quantum numbers
\begin{align}
{\rm totdeg}_{\boldsymbol{\mathcal{E}}}({\boldsymbol{\Omega}})=1\ ,\qquad {\rm totfer}_{\boldsymbol{\mathcal{E}}}({\boldsymbol{\Omega}})=0\ ,
\end{align}
obeying the flatness condition
\begin{align}
d{{\mathbf\Omega}}+{\mathbf\Omega}\wedge {\mathbf\Omega}\approx 0\ .
\end{align}
Using the decomposition of $\mathfrak{g}$ in Section \ref{Sec:2}, we split 
\begin{align}\label{superconformalconnection}
{\boldsymbol{\Omega}}={\boldsymbol{\Omega}}_{\mathfrak{h}}+\boldsymbol{E}_{+}+\boldsymbol{E}_{-}\ ,
\end{align}
where 
\begin{align}
\boldsymbol{\Omega}_{\mathfrak{h}}\in \Omega_{[0]}(\mathcal{M})\otimes \Omega(\boldsymbol{M})\otimes \mathfrak{h}
\end{align}
is a structure-group connection, and 
\begin{align}
\boldsymbol{E}_{\pm}\in \Omega_{[0]}(\mathcal{M})\otimes \Omega(\boldsymbol{M})\otimes (\mathfrak{g}_{\pm}/\mathfrak{h})
\end{align}
are sections. Using the $\mathfrak{h}$-covariant basis of Section \ref{Sec:2} and the spinor conventions in Appendix A, we expand 
\begin{align}\label{Omegadef}
{\boldsymbol{\Omega}}_{\mathfrak{h}} {}&:=i\left(\Omega_\beta{}^\alpha M_\alpha{}^\beta+ \sigma D + \Omega^{IJ}N_{IJ}\right)\ ,\\ \label{Eplusdef}
\boldsymbol{E}_{+}{}&:=i\left(E^{\alpha\beta}T_{\alpha\beta}+ F^{\alpha I} Q_{\alpha I}\right)\ ,\\ \label{Eminusdef}
\boldsymbol{E}_{-}{}&:=i\left(\widetilde{E}_{\alpha\beta} K^{\alpha\beta} +\widetilde F_{\alpha}{}^I S^\alpha{}_I \right)\ ,
\end{align}
with components from  $\Omega_{[0]}(\mathcal{M})\otimes \Omega(\boldsymbol{M})$ with AKSZ degrees, fermion parities, Grassmann--Koszul parities, and conformal weights as in Table \ref{tab:1}.
The $G_{\mathfrak{h}}$-covariant derivative
\begin{align}\label{superconformalnabla}
\nabla:=d+\Omega_{\mathfrak{h}}\ ,
\end{align}
acts on $\mathfrak{h}$-modules in $\boldsymbol{\mathcal{E}}(\boldsymbol{M})$ generated by elements from $\boldsymbol{\mathcal{A}}$ with components given by forms in $\Omega_{[0]}(\mathcal{M})\otimes \Omega(\boldsymbol{M})$. 
To differentiate components covariantly, we assume that the basis $\tau_i$ of $\boldsymbol{\mathcal{A}}$ is generated by taking direct products of a set of elements $\tau^\alpha,\tau_\alpha,\tau_I$ spanning the fundamental representations of $G_{\mathfrak{h}}$ obeying 
\begin{align}
\begin{aligned}
M_\alpha{}^\beta \tau_\gamma{}&=i\delta_\gamma^\beta \tau_\alpha-\frac{i}{4}\,\d_\a^\b \tau_\gamma\ ,\\
M_\alpha{}^\beta \tau^\gamma{}&=-i\delta_\alpha^\gamma \tau^\beta+\frac{i}{4}\,\d_\a^\b \tau^\gamma\ ,\\
N_{IJ} \tau_K{}&=-2i\eta_{K(J}\tau_{I)}\ ,
\end{aligned}
\qquad 
\begin{aligned}
D\tau_\alpha&=i\Delta^{(+)}_{\boldsymbol{\mathcal{A}}} \tau_\alpha\ ,\\
D\tau^\alpha&=i\Delta^{(-)}_{\boldsymbol{\mathcal{A}}} \tau^\alpha\ , \\
D\tau_I&=i\Delta^{(R)}_{\boldsymbol{\mathcal{A}}} \tau_I\ .  
\end{aligned}
\end{align}
Thus, letting 
\begin{align}
\psi^{(+)}:= f^\alpha \tau_\alpha\ ,\qquad \psi^{(-)}:= g_\alpha \tau^\alpha\ ,\qquad \psi^{(R)}:= h^I \tau_I
\end{align}
be sections, 
and defining 
\begin{align}
\nabla f^\alpha:= \tau^{\ast\alpha}(\nabla\psi^{(+)})\ ,\qquad 
\nabla g_\alpha:= \tau^{\ast}_{\alpha}(\nabla\psi^{(-)})\ ,\qquad
\nabla g^I:= \tau^{\ast I}(\nabla\psi^{(R)})\ ,
\end{align}
using dual basis elements obeying
\begin{align}
\tau^{\ast\alpha}(\tau_\beta)=\delta^\alpha_\beta\ ,\qquad
\tau^{\ast}_{\alpha}(\tau^\beta)=\delta^\beta_\alpha\ ,\qquad
\tau^{\ast I}(\tau_J)=\delta^I_J\ ,
\end{align}
one has 
\begin{align}\label{poincarenabla}
\nabla f^\alpha{}&= df^\alpha -\Omega^\alpha{}_\beta\wedge f^\beta -\Delta(f)\sigma\wedge  f^\alpha \ ,\\
\nabla g_\alpha{}&:=dg_\alpha+\Omega_\alpha{}^\beta \wedge g_\beta-\Delta(g)\sigma\wedge g_\alpha \ ,\\
\nabla h^I{}&:=dh^I+2\Omega^{IJ}\wedge  h_J-\Delta(h)\sigma\wedge h_I \ ,
\end{align}
using \eqref{A.26}. 
In particular, one has 
\begin{align}\label{flatpoincare}
\nabla E^{\alpha\beta} &=  dE^{\alpha\beta} + E^{\alpha\gamma}\wedge\Omega_\gamma{}^\beta- E^{\beta\gamma}\wedge\Omega_\gamma{}^\alpha -\s\wedge E^{\a\b}\ ,\\
\nabla F^{I\alpha}&= dF^{I\alpha}+ F^{I\gamma}\wedge\Omega_\gamma{}^\alpha+2\eta_{KJ}\O^{IJ}\wedge F^{\a K}-\frac12 \s\wedge F^{\a I}\ , \\
\nabla \widetilde E_{\alpha\beta} &=  d\widetilde E^{\alpha\beta} + \Omega_\a{}^\gamma\wedge \widetilde E_{\gamma\beta}- \Omega_\beta{}^\gamma\wedge \widetilde E_{\gamma\a} +\s\wedge \widetilde E_{\a\b}\ ,\\
\nabla \widetilde F^I_\alpha &= d\widetilde F^{I}_{\alpha}+ \Omega_\a{}^\beta\wedge \widetilde F_\beta^I +2\eta_{KJ}\O^{IJ}\wedge \widetilde F_\a^K+\frac12 \s\wedge \widetilde F_\a^I\ .
\end{align}
The flatness condition on ${\boldsymbol{\Omega}}$ amounts to 
\begin{align}\label{flatsuperconformal}
(R_M)_\beta{}^\alpha {}&\approx 0\ ,\qquad 
R_D \approx 0\ ,\qquad (R_N)^{IJ}\approx 0\ ,\\
\nabla E^{\alpha\beta} {}&\approx \frac{i}{2}\eta_{IJ}F^{\alpha I}\wedge F^{\beta J}\ ,\qquad \nabla F^{\alpha I} \approx -2 E^{\alpha\beta}\wedge \widetilde F_\beta{}^I\ ,\\
\nabla \widetilde E_{\alpha\beta}{}& \approx  \frac{i}{2}\eta_{IJ}\widetilde F_\alpha{}^I\wedge\widetilde F_\beta{}^J\ ,\qquad
\nabla \widetilde F_\alpha{}^I \approx 2 \widetilde E_{\alpha\beta}\wedge F^{\beta I}\ ,
\end{align}
where the superconformally covariantized curvatures 
\begin{align}
(R_M)_\beta{}^\alpha{}&:= (R_\Omega)_\beta{}^\alpha -4 E^{\alpha\gamma}\wedge \widetilde E_{\gamma\beta} + \eta_{IJ}F^{\alpha I}\wedge\widetilde F_\beta{}^J \\
  &+\d_\b^\a E^{\d\gamma}\wedge \widetilde E_{\gamma\d}-\frac{i}4 \d_\b^\a\,\eta_{IJ}F^{\gamma I}\wedge\widetilde F_\gamma{}^J\ ,\\
R_D {}& := R_\sigma+\frac{1}{2}\eta_{IJ} F^{\alpha I}\wedge\widetilde F_\alpha{}^J - 2i E^{\alpha\beta}\wedge\widetilde E_{\alpha\beta}\ ,\\
(R_N)^{IJ}{}&:=(R_\Omega)^{IJ}+F^{\alpha (I}\wedge\widetilde F_\alpha{}^{J)}\ ,
\end{align}
and the structure-group curvatures 
\begin{align}
(R_\Omega)_\beta{}^\alpha{}&:= d\Omega_\beta{}^\alpha +\Omega_\beta{}^\gamma\wedge\Omega_\gamma{}^\alpha-\Omega_\gamma{}^\alpha\wedge\Omega_\beta{}^\gamma\ ,\\
R_\sigma{}&:=d\sigma\ ,\\
(R_\Omega)^{IJ}{}&:=d\Omega^{IJ} +\Omega^{I}{}_L\wedge\Omega^{LJ}\ .
\end{align}

\subsection{Super-Poincar\'e backgrounds}

Imposing 
\begin{align}
\label{poincareconnection}
\boldsymbol{E}_{\mp}=0\ ,
\end{align}
which thus breaks $\pi_\sigma$, yields a background 
\begin{align}
\boldsymbol{\Omega}_{\pm}:=\boldsymbol{\Omega}_{\mathfrak{h}}+\boldsymbol{E}_{\pm}\in \Omega_{[0]}(\mathcal{M})\otimes \Omega(\boldsymbol{M})\otimes \mathfrak{g}_{\pm}\ ,
\end{align}
obeying 
\begin{align}\label{4.33}
R^{\boldsymbol{\Omega}_+}:= d\boldsymbol{\Omega}_++\boldsymbol{\Omega}_+ \wedge \boldsymbol{\Omega}_+\approx 0\ ,
\end{align}
decomposing into 
\begin{align}
(R_M)_\beta{}^\alpha =(R_\Omega)_\beta{}^\alpha{}\approx 0\ ,\qquad 
R_D = R_\sigma \approx 0\ ,\qquad (R_N)^{IJ}=(R_N)^{IJ}\approx 0\ .
\end{align}

\section{Linearized Two-form in Super-Poincar\'e Background}
\label{poincarebackground}

This section describes how the supersingleton arises as the chiral zero-form module of an abelian two-form $\boldsymbol{b}$ under the assumptions that i) the structure group is $G_{\mathfrak{h}}$; ii) the source is a (super)manifold $\boldsymbol{M}$ equipped with a flat connection $\boldsymbol{\Omega}_{+}$ valued in the super-Poincar\'e algebra $\mathfrak{g}_{+}$, viz., 
\begin{align}\label{5.27}
\nabla E^{\alpha\beta} \approx \frac{i}{2}\eta^{IJ}F_I^{\alpha}\wedge F_J^{\beta}\ ,\qquad \nabla F^{I\alpha}\approx 0\ ;
\end{align}
and iii) the two-form $\boldsymbol{b}$ is glued\footnote{A necessary condition for the cocycle to exist is that the super-Casimirs of the representations containing the two-form and zero-form are equal; indeed, one can verify that both vanish.} to a zero-form $\boldsymbol{j}$ with second-quantized conformal weights bounded from below by $+2$, viz.,
\begin{align}\label{5.1}
\Delta_{\Omega_{[0]}(\mathcal{M})}\left(\boldsymbol{j}\right)\stackrel{!}{\geqslant} 2\ ,
\end{align}
via a universally Cartan integrable cocycle $\boldsymbol{\Sigma}_+$ built from the background one-form $\boldsymbol{E}_{+}$ valued in $\mathfrak{g}_{+}/\mathfrak{h}$ (without any superframe assumption).
The main result is that (iii) implies
\begin{align}\label{5.3}
\boldsymbol{j}\stackrel{!}{=}\boldsymbol{h}^{(-)}\in  \Omega_{[0]}(\mathcal{M})\otimes \Omega(\boldsymbol{M})\otimes \mathsf{T}^{(-)}\ ,
\end{align}
where $\mathsf{T}^{(-)}$ is the $\mathfrak{g}_{+}$-module introduced in Section 3, whose completion into the supersingleton $\mathsf{S}$, which is a unitary, projective $G$-module, is left for future work; for further comments, see the Conclusions.

\subsection{Three-form cocycle}
\label{Subsection3formcocycle}
\paragraph{Basic set-up.}

The abelian two-form arises as the component in ghost number zero of an AKSZ superform ${\mathbf b}$ of total degree two and fermion parity zero belonging to the trivial representation of $\mathfrak{g}$, viz., 
\begin{align}
\boldsymbol{b}\in \Omega_{[0]}(\mathcal{M})\otimes \Omega(\boldsymbol{M})\otimes {\rm Id}_{\boldsymbol{\mathcal{A}}}\ ,
\end{align}
which thus has $\mathbb{Z}$-degree, fermion-parity, and conformal weight as follows:
\begin{align}
{\rm deg}_{\Omega_{[0]}(\mathcal{M})\otimes \Omega(\boldsymbol{M})}({\mathbf b})=2\ ,\qquad {\rm fer}_{\Omega_{[0]}(\mathcal{M})\otimes \Omega(\boldsymbol{M})}({\mathbf b})=0\ ,\qquad \Delta_{\Omega_{[0]}(\mathcal{M})}({\mathbf b})=0\ .
\end{align}
The on-shell deformation of the abelian curvature $d\boldsymbol{b}$ is described by a Cartan integrable system of constraints given by \eqref{4.33} and
\begin{align}\label{Rselfdual20}
\boldsymbol{R}^{\boldsymbol{b}}:= d{\boldsymbol{b} }+{\boldsymbol\Sigma}_+(\boldsymbol{{j}})\approx 0\ ,\qquad \boldsymbol{R}^{\boldsymbol{j}}:=d \boldsymbol{j}+\boldsymbol{\Omega}_+\boldsymbol{j}\approx 0\ ,
\end{align}
where 
\begin{align}
\boldsymbol{j} \in  \Omega_{[0]}(\mathcal{M})\otimes \Omega(\boldsymbol{M})\otimes \mathsf{V}
\end{align}
is an AKSZ superform of total degree zero and fermion parity zero belonging to a $G_{\mathfrak{h}}$-tensorial $\mathfrak{g}_+$-module ${\mathsf{V}}$, and ${\boldsymbol\Sigma}_+$ is a cocycle map trilinear in $\boldsymbol{E}_{+}$, i.e., $
{\boldsymbol\Sigma}_+(\boldsymbol{{j}})\equiv {\boldsymbol\Sigma}_+(\boldsymbol{{E}}_+,\boldsymbol{{E}}_+,\boldsymbol{{E}}_+;\boldsymbol{{j}})$,
as a multi-linear map.
The constraint on $\boldsymbol{R}^{\boldsymbol{\Omega}_+}$ is universally integrable.
The universal integrability of $\boldsymbol{R}^{\boldsymbol{b}}\approx 0$ is equivalent to
\begin{align}\label{5.13}
\left( \nabla \boldsymbol{\Sigma}_+\right)(\boldsymbol{j})-\boldsymbol{\Sigma}_+\left(\nabla\boldsymbol{j}\right) \approx 0\ ,
\end{align}
where thus $\left( \nabla \boldsymbol{\Sigma}_+\right)(\boldsymbol{j})\equiv 3{\boldsymbol\Sigma}_+(\nabla \boldsymbol{{E}}_+,\boldsymbol{{E}}_+,\boldsymbol{{E}}_+;\boldsymbol{{j}})$.
Finally, the universal integrability of the constraint on $\boldsymbol{R}^{\boldsymbol{j}}$ is equivalent to specifying the $\mathfrak{g}_+$-module $\mathsf{V}$.

\paragraph{Cocycle components.}
Introducing the three-form building blocks
\begin{align}
\Sigma^{abc}&:=E^a\wdg E^b\wdg E^c\ ,\qquad \Sigma^{ab,\alpha I}:=E^a\wdg E^b\wdg F^{\alpha I}\ ,\\
\Sigma^{a,\alpha I,\beta J}&:=E^a\wdg F^{\alpha I}\wdg F^{\beta J}\ ,\qquad 
\Sigma^{\alpha I,\beta J,\gamma K}:= F^{\alpha I}\wdg F^{\beta J}\wdg F^{\gamma K}\ ,
\end{align}
where $\Sigma^{a,\alpha I,\beta J}$ and $\Sigma^{\alpha I,\beta J,\gamma K}$ are symmetric under exchanges of pairs of spinor and R-symmetry indices, one may thus expand
\begin{align}\label{cocycleH}
{\boldsymbol\Sigma}_+(\boldsymbol{{j}})=\frac{1}{3!}\Sigma^{abc}H_{abc}+\frac{1}{2!}\Sigma^{ab,\alpha I}\Psi_{ab,\alpha I}+\frac{1}{2!}\Sigma^{a,\alpha I,\beta J}\Phi_{a,\alpha I,\beta J}+\frac{1}{3!}\Sigma^{\alpha I,\beta J,\gamma K} J_{\alpha I,\beta J,\gamma K}\ ,
\end{align}
using components $(H_{abc},\Psi_{ab,\alpha I},\Phi_{a,\alpha I,\beta J },J_{\alpha I,\beta J,\gamma K}) \in \Omega_{[0]}(\mathcal{M})\otimes \Omega(\boldsymbol{M})$ with $\mathbb{Z}$-degrees, fermion-parities, and conformal weights as follows:
\begin{align}
{\rm fer}_{\Omega_{[0]}(\mathcal{M})\otimes \Omega(\boldsymbol{M})}(H_{abc},\Psi_{ab,\alpha I},\Phi_{a,\alpha I,\beta J },J_{\alpha I,\beta J,\gamma K}){}&=(0,1,0,1)\ ,\\
{\rm deg}_{\Omega_{[0]}(\mathcal{M})\otimes \Omega(\boldsymbol{M})}(H_{abc},\Psi_{ab,\alpha I},\Phi_{a,\alpha I,\beta J },J_{\alpha I,\beta J,\gamma K}){}&=(0,0,0,0)\ ,\\
\Delta_{\Omega_{[0]}(\mathcal{M})}(H_{abc},\Psi_{ab,\alpha I},\Phi_{a,\alpha I,\beta J },J_{\alpha I,\beta J,\gamma K}){}&=(3,5/2,2,3/2)\ .   
\end{align}

\paragraph{Main result.}
Imposing \eqref{5.1}, i.e.,   
\begin{align}
\Delta_{\boldsymbol{\mathcal{A}}}(\mathsf{V})\stackrel{!}{\leqslant} -2\ ,
\end{align}
implies that 
\begin{align}
 J_{\alpha I,\beta J,\gamma K}\stackrel{!}{=}0\ .
\end{align}
As we shall see in Section \ref{Sec:cocycle}, it follows that the integrability condition \eqref{5.13} implies i) the algebraic constraints 
\begin{align}\label{ac1}
\Phi_{a,\alpha I,\beta J}{}&=(\sigma_a)_{\alpha\beta} \Phi_{IJ}\ ,\qquad \eta^{IJ} \Phi_{IJ}=0\ ,\\
\Psi_{ab,\alpha I}&{}=(\sigma_{ab})_\alpha{}^\beta \Psi^{(-)}_{\beta I}\ ,\\\label{ac3}
H_{abc}{}&=H^{(-)}_{abc}\ ,
\end{align}
where $\Psi^{(-)}_{\beta I}$ is right-handed and $H^{(-)}_{abc}$ is anti-selfdual, i.e.,
\begin{align}
H^{(-)}_{abc}=(\tilde\sigma_{abc})^{\alpha\beta} H_{\alpha\beta}\ ;
\end{align}
and ii) the differential constraints
\begin{align}\label{1stIntgrabilityPoincare}
\nabla \Phi_{IJ}\approx{}& E^{\alpha\beta} \Phi^E_{\alpha,\beta ;IJ} - F^{\alpha K}\left(\frac{i}{4}\Psi_{\alpha K}\eta_{IJ}+i\Psi_{\alpha [I}\eta_{J]K}\right)\ ,\\ \label{2ndIntgrabilityPoincare}
\nabla \Psi_{\alpha I}\approx{}& E^{\beta\gamma} \Psi^E_{\alpha\beta,\gamma;I}+ F^{\beta J}\left(\Phi^E_{\alpha,\beta ;IJ}-\frac{i}{3}H_{\alpha\beta}\eta_{IJ}\right)\ ,\\ \label{3rdIntgrabilityPoincare}
\nabla H_{\alpha\beta}\approx {}&E^{\gamma\delta}H^E_{\alpha\beta\gamma,\delta}+F^{\gamma I} \left(-\Psi^E_{\alpha\beta,\gamma ;I}\right)\ ,
\end{align}
where $\Phi^E$, $\Psi^E$ and $H^E$ comprise the first level of Poincaré descendants in $\mathsf{V}$, 
and the terms linear in $F$ encode the Poincar\'e supersymmetry transformations of the primaries, i.e., the states with maximal conformal weight; for general formalism, see Appendix \ref{App:symmetriescocycles}.
Finally, as we shall see in Section \ref{zeroformmastersection}, it follows that 
\begin{align}
V\stackrel{!}{=}T^{(-)}\ ,
\end{align}
i.e., \eqref{5.3}, and the universally Cartan integrable on-shell system is hence given by
\begin{align}\label{5.25}
\boldsymbol{R}^{\boldsymbol{b}}:= d{\boldsymbol{b} }+{\boldsymbol\Sigma}_+\left(\boldsymbol{{h}}^{(-)}\right)\approx 0\ ,\qquad \boldsymbol{R}^{\boldsymbol{{h}}^{(-)}}:=d \boldsymbol{{h}}^{(-)}+\boldsymbol{\Omega}_+\boldsymbol{{h}}^{(-)}\approx 0\ ,
\end{align}
together with the flatness condition on $\boldsymbol{\Omega}_+$.

\subsection{Solving the cocycle integrability condition}\label{Sec:cocycle}

To derive the main result, the first step is to decompose the integrability condition Eq. \eqref{5.13} under the structure group, and derive the algebraic constraints \eqref{ac1}-\eqref{ac3}, and, thereafter, the differential constraints \eqref{1stIntgrabilityPoincare}-\eqref{3rdIntgrabilityPoincare}.

\paragraph{Derivation of algebraic constraints.} 

The $G_{\mathfrak{h}}$-covariance implies  
\begin{align}\label{5.18}
&\nabla\Phi_{a,\alpha I,\beta J}\approx E^{\gamma\delta} \Phi^E_{\gamma,\delta;a,\alpha I,\beta J,b} +F^{\gamma K}\Phi^F_{\gamma K;a,\alpha I,\beta J}\ ,\\\label{5.19}
&\nabla\Psi_{ab,\alpha I}\approx E^{\beta\gamma}\Psi^E_{\beta,\gamma;ab,\alpha I}+F^{\beta J}\Psi^F_{\beta J;ab,\alpha I}\ ,\\ \label{5.20}
&\nabla H_{abc}\approx E^{\alpha\beta}H^E_{\alpha,\beta;abc}+F^{\alpha I}H^F_{\alpha I;abc}\ ,
\end{align}
which together with \eqref{5.27} imply that the integrability condition \eqref{5.13} splits into five linear, algebraic equations on the components in \eqref{cocycleH} and \eqref{5.18}--\eqref{5.20}, arising as the coefficients of $E^{\wedge k}\wedge F^{\wedge (4-k)}$, $k=0,1,\dots, 4$, to be solved in the order of increasing value of $k$. 
To this end, a useful lemma for further decomposition of $F^{\wedge (4-k)}$ is 
\begin{align}
(W_1\otimes W_2)^{\vee  r}=\bigoplus_{\vec \tau} {\rm Pr}_{\vec\tau}(W_1^{\otimes r})\otimes {\rm Pr}_{\vec\tau}(W_2^{\otimes r})\ ,
\end{align}
where $W_1$ and $W_2$ are vector spaces taken, respectively, as the ${\rm Spin}(1,5)$-quartet and $USp(4)$-quartet, and ${\rm Pr}_{\vec\tau}$ denote Young projectors of highest weight $\vec\tau$, each appears with multiplicity one.

\noindent\underline{\emph{$F^{\wedge 4}$-terms}}: This projection of the integrability condition reads
\begin{align}\label{F4}
F^{\alpha I}\wedge F^{\beta}{}_I \wedge F^{\gamma J}\wedge F^{\delta K} (\sigma^a)_{\alpha\beta} \Phi_{a,\gamma J,\delta K}\approx 0\ ,
\end{align}
where
\begin{align}
F^{\alpha I}\wedge F^{\beta}{}_I \wedge F^{\gamma J}\wedge F^{\delta K}
\in ({\rm Id}\otimes {\rm Tr})(\yngBlue{;}\otimes \yngRed{;})^{\vee 4}\ ,
\end{align}
which expands into
\begin{align}
F^{\alpha I}\wedge F^{\beta}{}_I \wedge F^{\gamma J}\wedge F^{\delta K}
\in\yngBlue{;;;,;}\otimes \yngRed{;;}\ \oplus\ \yngBlue{;;,;;}\otimes \yngRed{;,;}\ \oplus\  \yngBlue{;;,;,;}\otimes \yngRed{;;}\ \oplus\ 
\yngBlue{;;,;,;}\otimes \yngRed{;,;}\ \oplus\ 
\blueBullet \otimes\redBullet\ ,
\end{align}
using $\blueBullet$ and $\redBullet$ to denote singlets, and 
\begin{align}
(\sigma^a)_{\alpha\beta}\Phi_{a,\gamma J,\delta K}
\in\left(\yngBlue{;,;}\otimes \redBullet \right)\otimes (\yngBlue{;}\otimes \yngRed{;})^{\vee 2}\ ,
\end{align}
since  $\Phi_{a,\gamma J,\delta K}=\Phi_{a,\delta K,\gamma J}$, which expands into
\begin{align}
(\sigma^a)_{\alpha\beta}\Phi_{a,\gamma J,\delta K}
\in \yngBlue{;;;,;}\otimes \yngRed{;;}\ \oplus\ 
\yngBlue{;;,;;}\otimes \yngRed{;,;}\ \oplus\ 
\yngBlue{;;,;,;}\otimes \yngRed{;;}\ \oplus\ 
\yngBlue{;;,;,;}\otimes \yngRed{;,;}\ \oplus\ 
\blueBullet\otimes \yngRed{;,;}\ .
\end{align}
Thus, from \eqref{F4} it follows that
\begin{align}
\Phi \in \blueBullet\otimes \yngRed{;,;}_{\eta}\ ,
\end{align}
using an $\eta$-subscript  to denote a traceless R-symmetry tensor.
As a consequence, 
\begin{align}
\Sigma^{a,\alpha I,\beta J}\Phi_{a,\alpha I,\beta J}=(E\wedge F^{\wedge 2})^{IJ}\Phi_{IJ}\ ,\qquad \eta^{IJ}\Phi_{IJ}=0\ ,
\end{align}
and \eqref{5.18} collapses to
\begin{align}
\nabla \Phi_{IJ}\approx E^{\alpha\beta} \Phi^E_{\alpha,\beta; IJ} +F^{\alpha K}\Phi^F_{\alpha K;IJ}\ ,\qquad   \eta^{IJ}\Phi^E_{\alpha,\beta; IJ}=0\ ,\qquad \eta^{IJ}\Phi^F_{\alpha K;IJ}=0\ .
\end{align}

\noindent\underline{\emph{$E\wedge F^{\wedge 3}$-terms}}: Using the results obtained from the $k=0$ sector, the projection of the integrability condition to $k=1$ reads 
\begin{align}
-\frac1{2!} (E\wedge F^{\wedge 2})^{IJ} \wedge F^{\alpha K}\Phi^F_{\alpha K;IJ}-\frac{i}4 E^a \wedge F^{\alpha I}\wedge F^\beta{}_I \wedge F^{\gamma J} (\sigma^b)_{\alpha\beta} \Psi_{ab, \gamma J}\approx 0\ ,
\end{align}
where
\begin{align}
(E\wedge F^{\wedge 2})^{IJ} \wedge F^{\alpha K}{}&\in \left(\blueBullet\otimes \yngRed{;,;}\right)\otimes (\yngBlue{;}\otimes \yngRed{;})=\yngBlue{;}\otimes \yngRed{;;,;}\ \oplus\ \yngBlue{;}\otimes \yngRed{;}\ , \\
\Phi^F_{\alpha K;IJ}{}&\in \left(\blueBullet\otimes \yngRed{;,;}_\eta\right)\otimes (\yngBlue{;}\otimes \yngRed{;})=\yngBlue{;}\otimes \yngRed{;;,;}_\eta\ \oplus\ \yngBlue{;}\otimes \yngRed{;}\ ;
\end{align}
likewise,
\begin{align}
{}&E^{\delta\varepsilon}\wedge F^{\alpha I}\wedge F^\beta{}_I \wedge F^{\gamma J}\\\in{}& \left(\yngBlue{;,;}\otimes \redBullet\right)\otimes ({\rm Id}\otimes {\rm Tr})(\yngBlue{;}\otimes \yngRed{;})^{\vee 3}\\
={}& \left(\yngBlue{;,;}\otimes \redBullet\right)\otimes \left( \yngBlue{;;,;}\otimes \yngRed{;}\ \oplus\  \yngBlue{;,;,;}\otimes \yngRed{;}\right)\\
={}& \yngBlue{;;;,;;}\otimes \yngRed{;}\ \oplus\  
\yngBlue{;;;,;,;}\otimes \yngRed{;}\ \oplus\  
2\left(\yngBlue{;;,;;,;}\otimes \yngRed{;}\right)\ \oplus\  
2\left(\yngBlue{;}\otimes \yngRed{;}\right)\ ,
\end{align}
from which it follows that
\begin{align}
E^{\delta\varepsilon}\wedge F^{\alpha I}\wedge F^\beta{}_I \wedge F^{\gamma J}- E^{\alpha\beta}\wedge F^{\delta I}\wedge F^\varepsilon{}_I \wedge F^{\gamma J}\in 
\yngBlue{;;;,;,;}\otimes \yngRed{;}\ \oplus\  
\yngBlue{;;,;;,;}\otimes \yngRed{;}\ \oplus\  
\yngBlue{;}\otimes \yngRed{;}\ ,
\end{align}
and
\begin{align}
(\sigma^a)_{\delta\varepsilon}(\sigma^b)_{\alpha\beta} \Psi_{ab, \gamma J}\in\left(\yngBlue{;;,;,;}\otimes\redBullet\right)\otimes (\yngBlue{;}\otimes\yngRed{;})
=\yngBlue{;;;,;,;}\otimes \yngRed{;}\ \oplus\  
\yngBlue{;;,;;,;}\otimes \yngRed{;}\ \oplus\  
\yngBlue{;}\otimes \yngRed{;}\ .
\end{align}
It follows that 
\begin{align}\label{5.39}
\Phi^F \propto \Psi\in \yngBlue{;}\otimes \yngRed{;}\ .
\end{align}
Thus,
\begin{align}
\Sigma^{ab,\alpha I}\Psi_{ab,\alpha I}=(E^{\wedge 2}\wedge F)^{\alpha I}\Psi_{\alpha I}\ ,
\end{align}
and  \eqref{5.19} collapses to
\begin{align}
\nabla \Psi_{\alpha I}\approx E^{\beta\gamma} \Psi^E_{\beta,\gamma; \alpha I} +F^{\beta J}\Psi^F_{\beta  J;\alpha I}\ .
\end{align}

\noindent\underline{\emph{$E^{\wedge 2}\wedge F^{\wedge 2}$-terms}}: Using the results obtained from the $k=0,1$ sectors, the projection of the integrability condition to $k=2$ reads
\begin{align}
&-\frac12(E\wedge F^{\wedge 2})^{IJ}\wedge E^{\alpha\beta} \Phi^E_{\alpha,\beta;IJ}- \frac12(E^{\wedge 2}\wedge  F)^{\alpha I}\wedge F^{\beta J} \Psi^F_{\beta J; \alpha I}\cr &+
 \frac{i}{4}F^{\alpha I}\wedge F^{\beta}{}_I \wedge E^b\wedge E^c (\sigma^a)_{\alpha\beta}H_{abc}\approx 0\ ,
\end{align}
where
\begin{align}
(E\wedge F^{\wedge 2})^{IJ}\wedge E^{\alpha\beta}\in \yngBlue{;,;}\otimes \yngRed{;,;}\ ,\qquad
\Phi^E_{\alpha,\beta;IJ}\in \yngBlue{;,;}\otimes \yngRed{;,;}_\eta\ ;
\end{align}
likewise 
\begin{align}
{}&(E^{\wedge 2}\wedge  F)^{\alpha I}\wedge F^{\beta J}= E^a\wedge E^b (\sigma_{ab})_\gamma{}^\alpha \wedge F^{\gamma I}\wedge F^{\beta J}\\
\in {}&\left( {\Yfillcolour{blue}\yngBlue{;,;,;}}\shiftleft{1pt}{\raisebox{5pt}{\yngBlue{;}}}\ \otimes \redBullet\right)\otimes \left( {\Yfillcolour{blue}(\yngBlue{;}}\otimes \yngRed{;})\vee (\yngBlue{;}\otimes \yngRed{;})\right)\\
={}&\left( {\Yfillcolour{blue}\yngBlue{;,;,;}}\shiftleft{1pt}{\raisebox{5pt}{\yngBlue{;}}}\ \otimes \redBullet\right)\otimes \left( {\Yfillcolour{blue}\yngBlue{;}}{\shiftleft{1pt}{\yngBlue{;}}}\ \otimes \yngRed{;;}\ \oplus\ 
\yngBlue{;,!\bluefill;}\otimes \yngRed{;,;}\right)
\\
={}& \yngBlue{;;}\otimes\yngRed{;;}\ \oplus\ 
\yngBlue{;,;}\otimes\yngRed{;;}\ \oplus\ 
\yngBlue{;;}\otimes\yngRed{;,;}\ \oplus\ 
\yngBlue{;,;}\otimes\yngRed{;,;}\ ,
\end{align}
where the color filled cells are contracted into an $\epsilon$-tensor of ${\rm Spin}(1,5)$, and
\begin{align}
\Psi^F_{\beta J; \alpha I}\in \yngBlue{;;}\otimes\yngRed{;;}\ \oplus\ 
\yngBlue{;,;}\otimes\yngRed{;;}\ \oplus\ 
\yngBlue{;;}\otimes\yngRed{;,;}\ \oplus\ 
\yngBlue{;,;}\otimes\yngRed{;,;}\ ;
\end{align}
finally, 
\begin{align}
F^{\alpha I}\wedge F^{\beta}{}_I \wedge E^b\wedge E^c (\sigma_{bc})_\gamma{}^\delta
{}&\in\left(\yngBlue{;,;}\otimes \redBullet\right)\otimes \left(\yngBlue{;;,;,;}\otimes \redBullet\right)
=\left(\yngBlue{;;;,;;,;}\oplus \yngBlue{;;}\oplus \yngBlue{;;,;;,;;}\oplus \yngBlue{;,;}\right)\otimes \redBullet\ ,\\ 
(\sigma^{bc})_\gamma{}^\delta(\sigma^a)_{\alpha\beta} H_{abc} {}&\in \left(\yngBlue{;;}\oplus \yngBlue{;;,;;,;;}\right)\otimes \redBullet\ ,
\end{align}
whose contraction contains both irreps of $H_{abc}$.
It follows that 
\begin{align}
H\in \yngBlue{;;}\otimes \redBullet\ ,\qquad \Psi^F \in \yngBlue{;;}\otimes \redBullet\ \oplus\ \yngBlue{;,;}\otimes \yngRed{;,;}_\eta\ ,
\end{align}
with
\begin{align}\label{5.52}
\left.\Psi^F\right|_{ \yngBlue{;;}\otimes \redBullet}\propto H\ ,\qquad \left.\Psi^F\right|_{\yngBlue{;,;}\otimes \yngRed{;,;}_\eta}\propto \Phi^E\ .
\end{align}

\paragraph{Intermediate summary.}
At this stage, the cocycle has been reduced to
\begin{align}
{\boldsymbol\Sigma}_+(\boldsymbol{{j}})=\frac12 \Sigma^{\alpha\beta}H_{\alpha\beta}+\Sigma^{\alpha I}\Psi_{\alpha I}+\frac12 \Sigma^{IJ}\Phi_{IJ}\ ,
\end{align}
where the components obey 
\begin{align}
&\nabla H_{\alpha\beta}\approx E^{\gamma\delta}H^E_{\alpha\beta;\gamma\delta}+F^{\gamma I}H^F_{\alpha\beta;\gamma I}\ ,\\
&\nabla\Psi_{\alpha I}\approx E^{\beta\gamma}\Psi^E_{\alpha I;\beta\gamma}+F^{\beta J}\Psi^F_{\alpha I;\beta J}\ ,\\ 
&\nabla\Phi_{IJ}\approx E^{\alpha\beta} \Phi^E_{IJ;\alpha\beta} +F^{\alpha K}\Phi^F_{IJ;\alpha K}\ ,
\end{align}
with $\Phi^F$ and $\Psi^F$, respectively, given by \eqref{5.39} and \eqref{5.52}, and the three-form building blocks
\begin{align}
\Sigma^{\alpha\beta}&:=\frac{1}{3}\Sigma^{abc}(\tilde\sigma_{abc})^{\alpha\beta}=-\frac{8}{3}E^{\alpha\gamma_1}\wedge E_{\gamma_1\gamma_2}\wedge E^{\gamma_2\beta},\\ 
    \Sigma^{\alpha I}&:=\frac{1}{2!}(\sigma_{ab})_\beta{}^\alpha \Sigma^{ab,\beta I}=-2E_{\beta\gamma}\wedge E^{\gamma\alpha}\wedge F^{\beta I},\\ 
    \Sigma^{IJ}&:= 
    (\sigma_a)_{\alpha\beta}\Sigma^{a,\alpha I,\beta J}-\frac{\eta^{IJ}}{4}\eta_{KL}(\sigma_a)_{\alpha\beta}\Sigma^{a,\alpha K,\beta L}\cr &=-2E_{\alpha\beta}\wedge F^{\alpha I}\wedge F^{\beta J}+\frac{\eta^{IJ}}{2}\eta_{KL}E_{\alpha\beta}\wedge F^{\alpha K}\wedge F^{\beta L} \ .
\end{align}
In a super-Poincaré background, these blocks obey
\begin{align}
&\nabla\Sigma^{\alpha\beta}\approx -\frac{2i}{3}\eta_{IJ}\Sigma^{(\alpha I}\wedge F^{\beta)J} , \\
    &\nabla \Sigma^{\alpha I}\approx -i\eta_{JK} \Sigma^{IJ}\wedge F^{\alpha K} ,\\
    &\nabla\Sigma^{IJ}\approx 0\ .
\end{align}
They also obey the following algebraic relations:
\begin{align}\label{1stalgebraic}
&\Sigma^{\alpha\beta}\wedge E^{\gamma\delta}=-\frac43\epsilon^{\gamma\delta\rho(\alpha}\Sigma_{\rho}{}^{\beta)}, \\ \label{2ndalgebraic}
    &\Sigma^{\alpha I}\wedge E^{\beta\gamma}=-\frac12 \Sigma^{\alpha[\beta}\wedge F^{\gamma]I}-\frac34\epsilon^{\alpha\beta\gamma\delta}\Sigma_{\delta\rho}\wedge F^{\rho I},\\ \label{3rdalgebraic}
    &\Sigma^{IJ}\wedge E^{\alpha\beta}= -2\Sigma^{[\alpha [I}\wedge F^{\beta] J]}+\frac{\eta^{IJ}}{2}\eta_{KL}\Sigma^{[\alpha K}\wedge F^{\beta] L}.
\end{align}
where
\begin{align}
\Sigma_{\alpha}{}^\beta:=\frac{1}{4!}(\sigma_{abcd})_\alpha{}^\beta E^a\wedge E^b\wedge E^c\wedge E^d\ .
\end{align} 

\paragraph{Derivation of differential constraints.} The integrability condition \eqref{5.13} now reads
\begin{align}\label{finalintegrability}
&\frac12\Sigma^{\alpha\beta}\wedge E^{\gamma\delta}H^{E}_{\alpha\beta;\gamma\delta} + \left(-\frac12\Sigma^{\alpha\beta}\wedge F^{\gamma I} H^{F}_{\alpha\beta;\gamma I}+ \Sigma^{\alpha I}\wedge E^{\beta\gamma}\Psi^{E}_{\alpha I;\beta\gamma} \right)\cr 
    & +\left(-\frac{i}{3}\eta_{IJ}\Sigma^{\alpha I}\wedge F^{\beta J} H_{\alpha\beta}-\Sigma^{\alpha I}\wedge F^{\beta J}\Psi^F_{\alpha I;\beta J}-\frac12\Sigma^{IJ}\wedge E^{\alpha\beta}\Phi^E_{IJ;\alpha\beta} \right) \cr
    & +\left(-\frac{i}{2}\Sigma^{IJ}\wedge F^{\alpha K}\Psi_{\alpha I}\eta_{JK}-\frac12\Sigma^{IJ}\wedge F^{\alpha K} \Phi^F_{IJ;\alpha K} \right)\approx 0\ .
\end{align}
Using \eqref{1stalgebraic}, one can see that the first term of \eqref{finalintegrability} vanishes iff $H^E_{\alpha\beta;\gamma\delta}$ is symmetric in either $\alpha\beta\gamma$ or $\alpha\beta\delta$. Since it would be trivially zero, it cannot be symmetric in all its indices. Therefore $H^E_{\alpha\beta\gamma,\delta}$ is symmetric in $\alpha\gamma\delta$ and $H^E_{(\alpha\beta\gamma,\delta)}=0$, i.e., it is projected according to the Young diagram $\yngBlue{;;;,;}\otimes\redBullet$. 
In the second set of terms in the first line of \eqref{finalintegrability}, we use \eqref{2ndalgebraic}. The vanishing of the resulting $\Sigma_{\delta\rho}$-term implies that $\Psi^E_{\alpha I;\beta\gamma}$ has to be symmetric in $\alpha\beta$ or $\alpha\gamma$ but not symmetric in all spinor indices, i.e., $\Psi^E_{\alpha\beta,\gamma;I}$ belongs to the Young diagram $\yngBlue{;;,;}\otimes\yngRed{;}$. It follows that 
\begin{align}
    H^F_{\alpha\beta;\gamma I}=-\Psi^E_{\alpha\beta,\gamma;I}\ .
\end{align}
Turning to the second line of \eqref{finalintegrability}, $\Phi^E_{IJ;\alpha\beta}$ already belongs to an irreducible representation of $\mathfrak{su}^\ast(4)$. Using the notation $\Phi^E_{\alpha,\beta;I,J}$ to encode the Young-diagram structure $\yngBlue{;,;}\otimes\yngRed{;,;}$, ot follows from \eqref{3rdalgebraic} that 
\begin{align}
    \Psi^F_{\alpha I;\beta J}= \Phi^E_{\alpha,\beta ; IJ} -\frac{i}{3}\eta_{IJ}H_{\alpha\beta}\ .
\end{align}
Finally, the the third line of \eqref{finalintegrability} vanishes iff 
\begin{align}
    \Phi^F_{IJ;\alpha K}=-\frac{i}{4}\eta_{IJ}\Psi_{\alpha K}-i\Psi_{\alpha [I}\eta_{J]K}\ ,
\end{align}
which is $\eta$-traceless in $IJ$, which thus concludes the derivation of the first part of the main result.

\subsection{Solving the zero-form integrability condition}
\label{zeroformmastersection}

As the second, final step of the proof of the main result, we need to establish the integrability of $\boldsymbol{R}^{\boldsymbol{j}}\approx 0$ by specifying the $\mathfrak{g}$-module $\mathsf{V}$.
The first level of constraints has been known for a long time, e.g., \cite{Grojean:1998zt}.

\paragraph{Howe dual algebra.}
The higher levels can be constructed explicitly by introducing auxiliary variables $(q^\alpha,\bar q^\alpha, \xi^I)$ with\footnote{The conjugate variables of the super-Heisenberg algebra can be taken to be 
$$
p_\alpha:=i\frac{\partial}{\partial\bar q^\alpha}\ ,\qquad \bar p_\alpha:=-i\frac{\partial}{\partial q^\alpha}\ ,\qquad \bar\xi_I:=-\frac{\partial}{\partial\xi^I}\ .
$$}
\begin{align}
{\rm fer}_{\boldsymbol{\mathcal{A}}}(q^\alpha,\bar q^\alpha, \xi^I)=(0,0,1)\ ,\qquad 
\Delta_{\boldsymbol{\mathcal{A}}}(q^\alpha,\bar q^\alpha, \xi^I)=(-1/2,-1/2,0)\ .
\end{align}
Introducing Howe-dual $\mathfrak{su}(2)$ generators 
\begin{align}
    &L_+:=-i\bar q^\alpha\frac{\partial}{\partial q^\alpha}+i\Box_\xi,\quad 
    L_-:=iq^\alpha \frac{\partial}{\partial\bar q^\alpha}-i\xi^2, \\
    &L_0:=-\frac12\left(q^\alpha \frac{\partial}{\partial q^\alpha}-\bar q^\alpha \frac{\partial}{\partial\bar q^\alpha}+\xi^I \frac{\partial}{\partial\xi^I}-2\right)\ ,
\end{align}
where 
\begin{align}
    \xi^2:=\frac{1}{2}\eta_{IJ}\xi^I\xi^J\ ,\qquad   
    \xi^4:=\frac{1}{4!}\epsilon_{IJKL}\xi^I\xi^J\xi^K\xi^L\ ,\qquad
    \Box_\xi:=-\frac{1}{2}\eta^{IJ}\frac{\partial}{\partial\xi^I} \frac{\partial}{\partial\xi^J}\ ,
\end{align}
which satisfy
\begin{align}
    \Box_\xi \xi^2=2,\quad \Box_\xi(\xi^2\xi^I)=\xi^I,\quad \Box_\xi \xi^4=\xi^2,\quad 
    \xi^2\xi^2=2\xi^4\ ,
\end{align}
hence
\begin{align}
    [L_0,L_\pm]=\pm L_\pm, \quad [L_+,L_-]=2L_0\ ,
\end{align}
the associative higher-spin algebra $\boldsymbol{\mathcal{HS}}(V)$ is realized as the space of differential operators in $(q^\alpha,\bar q^\alpha, \xi^I)$ stabilized by the Howe dual algebra, and  $\mathsf{T}^{(-)}$ as the kernel of the Howe dual algebra \footnote{More generally, the spin-$j$ representation of the Howe dual algebra consists of $2j+1$ copies of the supersingleton with top-spin component given by an R-symmetry singlet,  chiral field strength with $\mathfrak{so}(1,5)$ weight $(s_j,s_j,s_j)^{(-)}$ where $s_j=1+j$; for a treatise of these fields for integer spin, see \cite{Basile:2024raj}.} in the space of formal power-series in $(q^\alpha,\bar q^\alpha, \xi^I)$.
The space of bilinear differential operators provides a representation of $\mathfrak{g}$ in $T^{(-)}$; in particular,
the oscillator realization of the Poincar\'e momentum and supercharge operators read
\begin{align}
T_{\alpha\beta}:=\frac{\partial}{\partial\bar q^\alpha}\frac{\partial}{\partial q^\beta}-\frac{\partial}{\partial\bar q^\beta}\frac{\partial}{\partial q^\alpha}\ ,\qquad 
    Q_{\alpha I}:=\frac{\partial}{\partial\xi^I}\frac{\partial}{\partial\bar q^\alpha}-\xi_I\frac{\partial}{\partial q^\alpha}\ .
\end{align}

\paragraph{Identification of chiral zero-form.}
We can thus impose \eqref{5.3}, i.e., identify the zero-form $\boldsymbol{j}$ with the chiral zero-form $\boldsymbol{h}^{(-)}$ realized as a power-series in $(q^\alpha,\bar q^\alpha, \xi^I)$ obeying 
\begin{align}
L_0 \boldsymbol{h}^{(-)} = 0\ ,\qquad 
L_\pm\boldsymbol{h}^{(-)}=0\ .
\end{align}
It is easy to see that the $L_0$ constraint requires only monomials of the type, suppressing the indices, $(q)^a(\bar q)^b(\xi)^c$ with $a-b+c=2$. The other constraints are better understood with a few examples. The $H_{\alpha\beta}$ field strength may appear in $q^\alpha q^\beta$, $q^\alpha\bar q^\beta\xi^2$, $q^\beta\bar q^\alpha\xi^2$ and $\bar q^\alpha\bar q^\beta\xi^4$. So, in principle, four different fields appear at this level. However, the $L_+$ constraint relates all four of them and implies that $H_{\alpha\beta}$ is symmetric in $\alpha\beta$. $\Phi_{IJ}$ can only couple with $\xi^I\xi^J$ and $L_+$ (or $L_-$) constraint implies $\eta^{IJ}\Phi_{IJ}=0$. The last low-level example is $\Psi_{\alpha I}$. The possible monomials are $q^\alpha\xi^I$ and $\bar q^\alpha\xi^2\xi^I$. To summarize, at the lowest level, the zero-form master field is 
\begin{align}
{\boldsymbol{h}}&=\left(q^\alpha q^\beta+q^\alpha\bar q^\beta\xi^2+\bar q^\alpha \bar q^\beta \xi^4\right)H_{\alpha\beta}+\left(q^\alpha\xi^I+\bar q^\alpha\xi^2\xi^I\right)\Psi_{\alpha I}+\left(\xi^I\xi^J\right)\Phi_{IJ}+\cdots.
\end{align}
At a higher level, a good example is $H^E_{\alpha\beta\gamma,\delta}$. The possible monomials that couple to it are $q^\alpha q^\beta q^\gamma \bar q^\delta$, $q^\alpha q^\beta\bar q^\gamma\bar q^\delta\xi^2$, $q^\alpha \bar q^\beta\bar q^\gamma\bar q^\delta\xi^4$, and all other possible permutations of the indices $\alpha\beta\gamma\delta$. The $L_-$ constraint implies that $H^E_{\alpha\beta\gamma,\delta}\in \yngBlue{;;;,;}$ and that it appears in $\boldsymbol{h}$ as 
\begin{align}
    \left(q^\alpha q^\beta q^\gamma \bar q^\delta -\frac32\bar q^\alpha\bar q^\beta q^\gamma q^\delta\xi^2-\bar q^\alpha \bar q^\beta \bar q^\gamma q^\delta \xi^4\right) H^E_{\alpha\beta\gamma,\delta}, 
\end{align}
or other equivalent ways.

Thus, in summary, the generalized Cartan curvature
\begin{align}\label{PoincareCartanCurvature}
\boldsymbol{R}^{\boldsymbol{h}^{(-)}}&:=d \boldsymbol{h}^{(-)}+\boldsymbol{\Omega}_+\boldsymbol{h}^{(-)}=\nabla{\boldsymbol{h}}^{(-)}+{\boldsymbol{E}}_+{\boldsymbol{h}}^{(-)}\ ,
\end{align}
using the oscillator realization
\begin{align}\label{5.89}
\boldsymbol{E}_+=2iE^{\alpha\beta} \frac{\partial}{\partial\bar q^\alpha}\frac{\partial}{\partial q^\beta}+iF^{\alpha I}\left(\frac{\partial}{\partial\xi^I}\frac{\partial}{\partial\bar q^\alpha}-\xi_I\frac{\partial}{\partial q^\alpha} \right)\ ,
\end{align}
is super-Poincar\'e covariant.
It is straightforward to verify that \eqref{PoincareCartanCurvature} reproduces \eqref{1stIntgrabilityPoincare}, \eqref{2ndIntgrabilityPoincare}, and \eqref{3rdIntgrabilityPoincare} (up to normalization) and similar on-shell relations for all higher descendants, which thus completes the second part of the proof of the main result.

\section{Linearized Two-form in Superconformal Background}
\label{superconformalbackground}

In the previous section, we described the three-form cocycle $\boldsymbol{\Sigma}_+$ that glues the chiral zero-form valued in the supersingleton to an abelian two-form in a flat super-Poincaré background $\boldsymbol{\Omega}_+$.
In what follows, we show that $\boldsymbol{\Sigma}_+$ provides the cocycle also on superconformal backgrounds $\boldsymbol{\Omega}$.

\subsection{Three-form cocycle}

Letting $\boldsymbol{b}$ and $\boldsymbol{h}^{(-)}$ be the same AKSZ superfields as in Section \ref{poincarebackground}, the following set of curvature constraints are universally Cartan integrable:
\begin{align}
\boldsymbol{R}^{\boldsymbol{\Omega}}&:=d\boldsymbol{\Omega}+\boldsymbol{\Omega}\wedge \boldsymbol{\Omega}\approx 0\ ,\\
\boldsymbol{R}^{\boldsymbol{b}}&:=d\boldsymbol{b}+\boldsymbol{\Sigma}\left(\boldsymbol{h}^{(-)}\right)\approx 0\ ,\\\label{6.3}
\boldsymbol{R}^{\boldsymbol{h}^{(-)}}&:=d\boldsymbol{h}^{(-)}+\boldsymbol{\Omega} \,\boldsymbol{h}^{(-)}\approx 0\ ,
\end{align}
where the superconformal cocycle is equal to the super-Poincar\'e cocycle, i.e.,
\begin{align}
\boldsymbol{\Sigma}\left(\boldsymbol{h}^{(-)}\right)=\boldsymbol{\Sigma}_+\left(\boldsymbol{h}^{(-)}\right)\ .
\end{align}
Thus, besides the addition of $\boldsymbol{\Omega}_-$, the superconformally covariant unfolded formulation uses the same field content and couplings as the super-Poincar\'e covariant unfolded formulation; in particular, it is not possible to add terms to $\boldsymbol{\Sigma}$ depending on $\tilde E$ and $\tilde F$ because of conservation of dilation weight. 

\paragraph{Proof of integrability.}

In the full superconformal background, we have that
\begin{align}
&\nabla\Sigma^{\alpha\beta}\approx -\frac{2i}{3}\eta_{IJ}\Sigma^{(\alpha I}\wedge F^{\beta)J} , \\
    &\nabla \Sigma^{\alpha I}\approx -i\eta_{JK} \Sigma^{IJ}\wedge F^{\alpha K} + \frac{3}{2}\Sigma^{\alpha\beta}\wedge\widetilde F_\beta{}^I ,\\
    &\nabla\Sigma^{IJ}\approx  -4 \Sigma^{\alpha [I}\wedge\widetilde F_\alpha{}^{J]}\ ,
\end{align}
using the $G_{\mathfrak{h}}$-covariant derivative $\nabla$ defined in (\ref{superconformalnabla}). 
The integrability of the cocycle implies that 
\begin{align}\label{6.8}
\nabla \Phi_{IJ}\approx{}& E^{\alpha\beta} \Phi^E_{\alpha,\beta ;IJ} + F^{\alpha K}\left(\frac{i}{4}\Psi_{\alpha K}\eta_{IJ}+i\Psi_{\alpha [I}\eta_{J]K}\right)+ \widetilde F_\alpha{}^K\Phi^{\widetilde F\; \alpha}{}_{IJ;K}\ ,\\\label{6.9}
\nabla \Psi_{\alpha I}\approx{}& E^{\beta\gamma} \Psi^E_{\alpha\beta,\gamma;I}+ F^{\beta J}\left(\Phi^E_{\alpha,\beta ;IJ}+\frac{i}{3}H_{\alpha\beta}\eta_{IJ}\right)+\widetilde F_\beta{}^J\Psi^{\widetilde F\; \beta}{}_{\alpha I;J}\ ,\\\label{6.10}
\nabla H_{\alpha\beta}\approx {}&E^{\gamma\delta}H^E_{\alpha\beta\gamma,\delta}+F^{\gamma I} \left(-\Psi^E_{\alpha\beta,\gamma ;I}\right)+\widetilde F_\gamma{}^I H^{\widetilde F\; \gamma}{}_{\alpha\beta; I}\ ,
\end{align}
where $(\Phi^{\widetilde F\; \alpha}{}_{IJ;K},\Psi^{\widetilde F\; \beta}{}_{\alpha I;J},H^{\widetilde F\; \gamma}{}_{\alpha\beta; I})$ are to be found. Following the same steps as in the previous section, we obtain
\begin{align}\label{1stsuperconTransf}
    \Phi^{\widetilde F\; \alpha}{}_{IJ;K}&=0\ ,\\ \label{2ndsuperconfTransf}
    \Psi^{\widetilde F\; \beta}{}_{\alpha I;J} &= -4\delta^\beta_\alpha \Phi_{IJ}\ ,\\ \label{3rdsuperconfTransf}
    H^{\widetilde F\; \gamma}{}_{\alpha\beta; I}& =\frac{3}{2}\delta^\gamma_{(\alpha}\Psi_{\beta) I}\ .
\end{align}
Using the Lemma of Appendix \ref{App:symmetriescocycles}, the action of Killing supertranslations $\delta_Q\equiv \epsilon^{\alpha I}Q_{\alpha I}$ and superconformal transformations $\delta_S\equiv \widetilde\epsilon_\alpha{}^I S^\alpha{}_I$, with parameters $\epsilon^{\alpha I}$ and $\widetilde\epsilon_\alpha{}^I$ annihilating the background, on $(\Phi_{IJ},\Psi_{\alpha I},H_{\alpha\beta})$ can be read off from \eqref{6.8}-\eqref{6.10} with \eqref{1stsuperconTransf}-\eqref{3rdsuperconfTransf} imposed.
Thus, the absence of an $\widetilde F_\alpha{}^I$-term in the Cartan curvature for $\Phi_{IJ}$ confirms its status as a superconformal primary, and the  
$\widetilde F_\alpha{}^I$-terms in \eqref{6.9} and \eqref{6.10} provide the superconformal transformations of $\Psi_{\alpha I}$ and $H_{\alpha\beta}$, respectively.
As a further check, matching $F$- and $\widetilde F$-terms to supersymmetry and superconformal transformations, respectively, we find 
\begin{align}
    [\delta_Q,\delta_S]\Phi_{IJ} &= -\delta_S\delta_Q \Phi_{IJ} =
    -\delta_S\left( \frac{i}{4}\epsilon^{\alpha K}\Psi_{\alpha,K}\eta_{IJ}+i\epsilon^{\alpha K}\Psi_{\alpha,[I}\eta_{J]K}\right) \cr
    &=-i\epsilon^{\alpha K}\widetilde\epsilon_\alpha{}^L\Phi_{KL}\eta_{IJ}+ 4i\epsilon^{\alpha K}\widetilde\epsilon_\alpha{}^L\Phi_{L[I}\eta_{J]K} \cr 
    &= -i\epsilon^{\alpha [K}\widetilde\epsilon_\alpha{}^{L]}\left(\Phi_{KL}\eta_{IJ}+ 4\Phi_{[K|[I}\eta_{J]|L]}\right) + 4i\epsilon^{\alpha (K}\widetilde\epsilon_\alpha{}^{L)}\Phi_{K[I}\eta_{J]L}\ ,
\end{align}
where the first term can be rewritten using that 
$\Phi_{[IJ}\eta_{KL]}$ is proportional to $\epsilon_{IJKL}\Phi_{MN}\eta^{MN}=0$, which implies 
\begin{align}
    \Phi_{IJ}\eta_{KL}+4\Phi_{[I|[K}\eta_{L]|J]}+\Phi_{KL}\eta_{IJ}=0\ .
\end{align}
Thus,
\begin{align}
     [\delta_Q,\delta_S]\Phi_{IJ}=i\epsilon^{\alpha K} \widetilde\epsilon_{\alpha K}\Phi_{IJ} +4i \epsilon^{\alpha (K}\widetilde\epsilon_\alpha{}^{L)}\Phi_{K[I}\eta_{J]L}\ ,
\end{align}
in agreement with \eqref{QSanticommutator}, i.e., 
\begin{align}
[\delta_Q,\delta_S]=\epsilon^{\alpha I}\widetilde\epsilon_{\beta I} M_\alpha{}^\beta +\frac{1}{2}\epsilon^{\alpha I}\widetilde\epsilon_{\alpha I}D -\epsilon^{\alpha (I}\widetilde\epsilon_\alpha{}^{J)} N_{IJ}\ ,
\end{align}
provided that $\Phi_{IJ}$ has conformal dimension $2$.

\subsection{Superconformal chiral zero-form}

The oscillator realization equips the chiral zero-form module $T^{(-)}$ with a left $\mathfrak{g}$ action, ensuring the universal integrability of \eqref{6.3}.
More explicitly, 
\begin{align}\label{superconformalmastereq}
{\boldsymbol{R}}^{\boldsymbol{h}^{(-)}}=\nabla{\boldsymbol{h}}+{\boldsymbol{E}}_{+}{\boldsymbol{h}}^{(-)}+ {\boldsymbol{E}}_{-}{\boldsymbol{h}}^{(-)}\approx 0\ ,
\end{align}
where the oscillator realization of ${\boldsymbol{E}}_{+}$ is given in \eqref{5.89}, and 
\begin{align}
{\boldsymbol{E}}_{-}=\widetilde E_{\alpha\beta}\left( 2\bar q^\alpha q^\beta\right)+\widetilde F_\alpha{}^I\left( -\eta^{IJ}\frac{\partial}{\partial\xi^J}\bar q^\alpha-\xi^I q^\alpha\right)\ ,
\end{align}
as can be seen using the definition of  ${\boldsymbol{E}}_{-}$ in \eqref{Eminusdef} together with the following first-quantized realizations of the special conformal and superconformal generators: 
\begin{align}
K^{\alpha\beta}:=\bar q^\alpha q^\beta-\bar q^\beta q^\alpha,\quad 
    S^{\alpha I}:= -\eta^{IJ}\frac{\partial}{\partial\xi^J}\bar q^\alpha-\xi^I q^\alpha\ .
\end{align}
As in the super-Poincaré background, \eqref{superconformalmastereq} contains the on-shell equations for all descendants of $(H_{\alpha\beta},\Psi_{\alpha I},\Phi_{IJ})$ as well as their supersymmetry, superconformal and special conformal transformations, which are contained in the coefficients of $F^{\alpha I}$, $\widetilde F_\alpha{}^I$ and $\widetilde E_{\alpha\beta}$, respectively.

\section{Conclusion and Prospects}
\label{conclusion}

In this paper, we have provided an unfolded formulation of the six-dimensional $(2,0)$ tensor multiplet as a universally Cartan integrable system of constraints on a the exterior derivatives of a chiral zero-form $\boldsymbol{h}$, two-form potential $\boldsymbol{b}$, and connection one-form $\boldsymbol{\Omega}$, all of which can be thought of as horizontal forms on a family of correspondence spaces $\boldsymbol{C}$ making up components of a yet-to-be constructed fully non-linear superconnection $\boldsymbol{z}$ playing the role of fundamental field of an underlying AKSZ sigma model.
The one-form is treated as a background field valued in the superconformal algebra $\mathfrak{g}$, while the zero-form is treated as a linearized fluctuation valued in an infinite-dimensional, unitarizable representation of $\mathfrak{g}$.
The zero-form is glued to the two-form, which belongs to the singlet of $\mathfrak{g}$, via a cocycle whose integrability on general superconformal backgrounds is the main result of this work.
In particular, on a source of superdimension $(6|16)$ equipped with a background connection valued in the super-Poincaré subalgebra of the superconformal algebra with invertible superframe fields, the unfolded system yields the superspace formulation given originally in \cite{Howe:1983fr,Koller:1982cs}; see also \cite{Grojean:1998zt}.

We have focused on the $G_{\mathfrak{h}}$-tensorial unfolding of the tensor multiplet, leaving the harmonic analysis, i.e., the construction of linearized solution spaces with distinct boundary conditions, for future work \cite{paper0}.
We expect that starting from the projective, unitary supersingleton module $\mathsf{S}$ of the superconformal group $G$, Green's functions obeying various boundary conditions arise from elements in extended supersingletons $\mathsf{S}_\infty$ governed by a generalization of the metaplectic group underlying four-dimensional higher-spin gravity and three-dimensional conformal field theory \cite{fibre,meta,FSG1} to the current context.
More precisely, it is anticipated that $G$ extends into a holomorphic, complexified, metaplectic group ${\rm Mp}(G^{\mathbb{C}})$ with a proper oscillator realization in $\mathsf{S}$, whose asymptotic region contains projectors with separate left and right polarizations giving rise to the $G$-modules in  $\mathsf{S}_\infty$ glued together by complexified Bogolyubov transformations from the interior of $G^{\mathbb{C}}$.
In particular, we envisage elements $\sigma,\kappa \in {\rm Mp}(G^{\mathbb{C}})$ whose adjoint action yield the $\mathfrak{g}$-automorphisms $\pi_\sigma$ and $\pi_\kappa$ and corresponding $G$-automorphisms $\Pi_\sigma$ and $\Pi_\kappa$ introduced in Section 3.3 (their explicit realization is given in \cite{FSG1}), and whose one-sided actions connect the oscillator realizations of $\mathsf{W}^\pm$ and $\mathsf{D}^\pm$.

There are several directions to generalize the results presented here. One natural direction is the unfolded description of on-shell $(2,0)$ conformal supergravity \cite{Bergshoeff:1999db} whose supermultiplet contains five anti-selfdual tensors; this system can then be coupled to additional tensor multiplets appearing as matter. In general, the complete unfolding of systems containing gravity remains an open problem already in the case of pure gravity; for the terms of the Q-structure quadratic in the chirall zero-form, see \cite{Vasiliev:1989xz}, and for the full Cartan integration formula \eqref{C.23} with gauge parameters corresponding to normal coordinates, see \cite{Muller:1997zk}.
One may speculate that a sufficient amount of supersymmetry will provide enough constraints to derive a complete answer for the Cartan curvatures. 
To this end, one may seek to combine the exact unfolding of four-dimensional Yang--Mills theory, as well as of electromagnetism coupled to charged, scalar matter in Minkowski spacetime, given recently in \cite{Misuna:2024ccj,Misuna:2024dlx}, with the manifest conformal covariance of our approach, to unfold the prototypical superconformal field theory, namely four-dimensional, maximally supersymmetric Yang-Mills theory, whose ``square'' might then contain information about unfolded maximal supergravity.

Another direction concerns nonlinear, unfolded descriptions of coinciding M-theory five-branes. 
As a first step, one may consider coupling a two-form $B$ with anti-selfdual curvature to a conserved four-form current $J$. When $J=0$, $B$ is sourced by the anti-selfdual curvature zero-form $H^-_{abc}$ via a cocycle proportional $e^a \wedge e^b\wedge e^c H^-_{abc}$. The coupling of $B$ to $J$ can be achieved by introducing a three-form $C$, viz.,
\begin{align}
dB+C+\frac1{3!}e^a\wedge e^b\wedge e^c H^-_{a,b,c}\approx 0\ ,\qquad 
dC+\frac1{4!}e^a\wedge e^b\wedge e^c\wedge e^d J_{a,b,c,d}\approx 0\ ,\\
\nabla H^-_{a,b,c}+ e^d H^-_{ad,b,c}+ e^d J_{a,b,c,d}-3e_{[a} \ast J_{b,c]}\approx 0\ ,\qquad 
\nabla J_{a,b,c,d}+e^e J_{ae,b,c,d}\approx 0\ ,
\end{align}
and a tower of Cartan integrable constraints\footnote{The constraints on $\nabla H^-_{a(1+n),b,c}$ contain trace-projected source terms.} on the Lorentz-covariant exterior derivatives of the descendants $H^-_{a(1+n),b,c}$ and $J_{a(1+n),b,c,d}$ for $n=1,2,3,\dots$; we use conventions in which
\begin{align}
\ast T_{a[6-p]}=\frac{1}{p!}\epsilon_{a[6-p]b[p]}T^{b[p]}\ ,
\end{align}
from which it follows that $J_{a,b,c,d}=-\frac12 \epsilon_{abcdmn}\ast J^{mn}$, hence
\begin{align}
    \frac1{3!}\epsilon_{abc}{}^{mnp} e^d J_{m,n,p,d}=-\frac1{2\cdot 3!}\epsilon_{abc}{}^{mnp}\epsilon_{mnpdxy} e^d \ast J^{xy}=-3e_{[a}\ast J_{bc]}\ .
\end{align}
Notably, the AKSZ formalism treats $C$ as an independent fundamental field in the path integral, which relaxes the constraints on finding interactions as obtaining $J$ in the image of $d$ acting on (ultra-local) functionals is unnecessary.

To go further, one may be inspired by the embedding of conformal theories in three dimensions into flat, horizontal superconnections on noncommutative , associative correspondence spaces of the same type used in four-dimensional higher-spin gravity found recently in \cite{FSG1,FSG2} in non-supersymmetric cases.
We expect these to extend naturally to models consisting of $\mathfrak{osp}(8|4)$ supersingletons with self-interactions governed by noncommutative  rather than Lorentzian geometry. 
To this end, expecting on general grounds \cite{DouglasTalk} that the interacting, conformal five-brane theory consists of a finite set of tensor-multiplets, one may assemble their chiral zero-forms, $\boldsymbol{h}^i(q,\bar{q};\xi)$ say, into symbols $\boldsymbol{c}:=\boldsymbol{h}^i(q^\alpha,\bar{q};\xi)|(-2i)\rangle\langle e_i|$ of operators depending on the complete set of oscillators $(q,p,\bar q,\bar p,\xi,\bar\xi)$, carrying a left-action of $\boldsymbol{\mathcal{HS}}(V)
$ and a right-action of an internal algebra acting faithfully on $\langle e_i|$, and obeying 
\begin{align}
d\boldsymbol{c}+\boldsymbol{\Omega} \star \boldsymbol{c}\approx0\ ,\qquad d\boldsymbol{\Omega}+\boldsymbol{\Omega}\star \boldsymbol{\Omega}\approx 0\ ,
\end{align}
in linearization around superconformal vacua $\boldsymbol{\Omega}$.
One may then use a pair of closed and (anti-)holomorphic four-forms available in the correspondence space in question, to seek an embedding of the resulting vacuum system into a flat, horizontal superconnection valued in the direct product of ${\rm mat}_{1|1}$ and a generalization of the fractional-spin algebra used in the three-dimensional context, but now with representation content correlated to form-degree mod four instead of mod two, to trigger, at the first orders, linearized cocycles of the form constructed in this paper, and quadratic source terms for one-forms valued in the higher-spin algebra and the internal algebra built from $\boldsymbol{c}\star \bar{\boldsymbol{c}}$ and $\bar{\boldsymbol{c}}\star \boldsymbol{c}$, respectively. 
In the three-dimensional case, the latter provides nonlocal sources for the internal gauge fields which is how the construction escapes no-go theorems for self-interacting, three-dimensional supersingleton theories with $32$ supersymmetries \cite{Nicolai:1984gb}, providing the rationale why the approach sketched above may pave a way towards the self-interacting five-brane.

A key feature of the AKSZ sigma model introduced in Section 2, is that its partition function is a sum over sources $\boldsymbol{X}=T[1]\boldsymbol{M}$ with $\boldsymbol{M}$ of different (super)dimensions, inducing a quantization functor mapping classical integration modules to operator algebras.
In particular, the quantization of real symplectic leaves of the target $\boldsymbol{Y}$ takes place on sources $T[1]\boldsymbol{D}_2$ with $\boldsymbol{D}_2$ given by two-dimensional disks \cite{CaFe}. Applied to unfolded free fields on spacetime backgrounds, their creation and annihilation operators arise as noncommutative  coordinates of their chiral zero-form modules, and the spacetime coordinates as integration constants for coset elements in Cartan gauge groups which can be added as separate fundamental fields by extending the targets of the superconnections to Lie groupoids.
In particular, combined with algebraic structures of the first-quantized metaplectic group algebra, the formalism provides a path-integral approach towards quantizing  supersingletons in two, three, four and six dimensions that does not rely on any spacetime Lagrangian hence providing a natural approach to the chiral bosons in two and six dimensions.

Finally, one of the main motivations behind our work, is that it can be extended naturally to $\mathfrak{osp}(8^\ast|8)$, which allows us to study another elusive conformal theory in six dimensions: Hull's exotic $(4,0)$ gravity \cite{Hull:2000zn, Hull:2000rr}. Hull conjectured that if an interacting theory exists for the $(4,0)$ singleton multiplet, it is the strong coupling limit of $N=4$ supergravity in five dimensions. A recent overview and summary of the difficulties in formulating the interacting theory is \cite{Hull:2022vlv}. The construction of the multiplet using the oscillator method was given in \cite{Chiodaroli:2011pp}. Other recent developments are the construction of a free supersymmetric action for the multiplet breaking explicit $\mathfrak{so}(1,5)$ covariance \cite{Bertrand:2022pyi} and the derivation of the $(4,0)$ spectrum from two copies of the $(2,0)$ singleton using the heuristic idea gravity$=$(gauge theory)${}^2$ \cite{Borsten:2017jpt}. We hope the formalism presented here paves the way for constructing the complete interacting theory. Work on the free unfolded theory will appear in \cite{4comma0}.  

\paragraph{Acknowledgments.} We thank R. Aros, L. Avilés, T. Basile, N. Boulanger, F. Caro, F. Diaz, O. Fuentealba, M. Grigoriev, N. Merino, E. Skvortsov, D. Tiempo, and O. Valdivia, M. Valenzuela for useful discussions. 
P.S. expresses his gratitude to the CECs in Valdivia for hospitality during the initial stages of this work and acknowledges UNAP Consolida grant ``Higher-spin inspired IR modifications of 3d gravity'',; ANID grant Regular N°1250672; and Proyecto MATH-Sud for partial support.

\begin{appendix}

\section{Notation and Conventions} 
\label{notation}
We let 
\begin{itemize}
    \item  boldfaced, capital, Roman letters denote manifolds, e.g., sources and targets of AKSZ sigma models; 

    \item calligraphic, capital, Roman letters denote spaces of maps, e.g., configuration spaces of AKSZ sigma models; 

    \item boldfaced, calligraphic, capital, Roman letters denote associative dg-algebras; 

    \item minuscule, Gothic letters denote Lie (super)algebras;

    \item capital, Roman letters in sans serif font denote modules and representations of Lie and associative algebras; 

    \item boldfaced, minuscule, Roman and Greek letters denote superconnections valued in direct products of first- and second-quantized algebras; 

    \item capital, Roman letters denote components of superconnections expanded over first-quantized fiber algebras;

    \item hatted, minuscule, Roman and Greek letters denote generators of first-quantized fiber algebra; and

    \item hatted, capital, Roman letters denote basis elements of first-quantized fiber algebras.
    
    \item $(\cdots)\downarrow_{\mathfrak{a}}$ means decomposition with respect to $\mathfrak{a}$.
\end{itemize}

\paragraph{Gradings.} The targets $\boldsymbol{Y}$ are parity-shifts of differential, $\mathbb{Z}$-graded, associative superalgebras (DGA), $\boldsymbol{\mathcal{A}}$, i.e.,
\begin{align}
\boldsymbol{Y}=\boldsymbol{\mathcal{A}}[1]\ ,
\end{align}
in turn realized as composite operators built from the coordinates of fibre space $\boldsymbol{F}$, viz.,
\begin{align}
\boldsymbol{\mathcal{A}}\cong \Omega_{[0]}(\boldsymbol{F})\ ,
\end{align}
arising as the fibre manifold of the target of an auxiliary, first-quantized system.
This algebra is 
equipped with fermion-parity and $\mathbb{Z}$-degree maps
\begin{align}
{\rm fer}_{\boldsymbol{\mathcal{A}}}\in \{0,1\}  \ ,\qquad 
{\rm deg}_{\boldsymbol{\mathcal{A}}}\in \mathbb{Z}\ .
\end{align}
The sources are $\mathbb{Z}$-graded supermanifolds given by parity-shifted tangent bundles $\boldsymbol{X}=T[1]\boldsymbol{M}$ over supermanifolds $\boldsymbol{M}$, such that $\Omega(\boldsymbol{X})$ are bi-DGAs equipped with fermion-parity, $\mathbb{Z}$-degree and form-degree maps; projecting to zero form degree, there is an isomorphism $\Omega_{[0]}(\boldsymbol{X})\cong \Omega(\boldsymbol{M})$, the DGA of forms on $\boldsymbol{M}$, such that 
\begin{align}
{\rm fer}_{\Omega(\boldsymbol{M})}={\rm fer}_{\Omega_{[0]}(\boldsymbol{X})}\in \{0,1\}  \ ,\qquad 
{\rm deg}_{\Omega(\boldsymbol{M})}={\rm deg}_{\Omega_{[0]}(\boldsymbol{X})}\in \mathbb{Z}\ .
\end{align}
The configuration spaces $\mathcal{M}$ are $\mathbb{Z}$-graded supermanifolds, and $\Omega_{[0]}(\mathcal{M})$, the DGAs of functionals on $\mathcal{M}$, have fermion-parity and $\mathbb{Z}$-degree maps
\begin{align}
{\rm fer}_{\Omega_{[0]}(\mathcal{M})}\in \{0,1\}  \ ,\qquad 
{\rm gh}_{\Omega_{[0]}(\mathcal{M})}\equiv {\rm deg}_{\Omega_{[0]}(\mathcal{M})} \in \mathbb{Z}\ ,
\end{align}
where the $\mathbb{Z}$-degree map is referred to as the (second-quantized) ghost number.
Fields, gauge parameters, and equations of motion are assumed to be elements of the superbundle 
\begin{align}
\boldsymbol{\mathcal E}(\boldsymbol{X}):=\Omega_{[0]}(\mathcal M(\boldsymbol{X})\times \boldsymbol{X}\times \boldsymbol{F})\cong \Omega_{[0]}(\mathcal{M}(\boldsymbol{X}))\otimes \Omega_{[0]}(\boldsymbol{X})\otimes \boldsymbol{\mathcal{A}}\ ,   
\end{align}
or its isomorphic version
\begin{align}
\boldsymbol{\mathcal E}(\boldsymbol{M}):=\Omega_{[0]}(\mathcal M(\boldsymbol{M}))\otimes \Omega(\boldsymbol{M})\otimes  \boldsymbol{\mathcal{A}}\ ,   
\end{align}
equipped with triplets of fermion parities and $\mathbb{Z}$-degrees, viz.,
\begin{align}
{\rm fer}_{\boldsymbol{\mathcal E}(\boldsymbol{M})}{}&=({\rm fer}_{\Omega_{[0]}(\mathcal{M})},{\rm fer}_{\Omega(\boldsymbol{M})},{\rm fer}_{\boldsymbol{\mathcal{A}}})\in \{0,1\}\times \{0,1\}\times \{0,1\}\ ,\\
{\rm deg}_{\boldsymbol{\mathcal E}(\boldsymbol{M})}{}&=({\rm deg}_{\Omega_{[0]}(\mathcal{M})},{\rm deg}_{\Omega(\boldsymbol{M})},{\rm deg}_{\boldsymbol{\mathcal{A}}})\in \mathbb{Z}\times \mathbb{Z}\times\mathbb{Z}\ ,
\end{align}
idem ${\rm fer}_{\boldsymbol{\mathcal E}(\boldsymbol{M})}$ and ${\rm deg}_{\boldsymbol{\mathcal E}(\boldsymbol{M})}$.
The total fermion parity and AKSZ $\mathbb{Z}$-degree are defined by
\begin{align}\label{1.7}
{\rm totfer}_{\boldsymbol{\mathcal E}(\boldsymbol{M})}{}&:={\rm fer}_{\Omega_{[0]}(\mathcal{M})}+{\rm fer}_{\Omega(\boldsymbol{M})}+{\rm fer}_{\boldsymbol{\mathcal{A}}}\ {\rm mod}\ 2\ ,\\
\label{1.8}{\rm totdeg}_{\boldsymbol{\mathcal E}(\boldsymbol{M})}{}&:={\rm deg}_{\Omega_{[0]}(\mathcal{M})}+{\rm deg}_{\Omega(\boldsymbol{M})}+{\rm deg}_{\boldsymbol{\mathcal{A}}}\ ,
\end{align}
idem ${\rm totfer}_{\boldsymbol{\mathcal E}(\boldsymbol{M})}$ and ${\rm totdeg}_{\boldsymbol{\mathcal E}(\boldsymbol{M})}$.
The Grassmann bi-parity
\begin{align}\label{1.9b}
{\rm Gr}_{\boldsymbol{\mathcal E}}=({\rm totfer}_{\boldsymbol{\mathcal E}},{\rm totdeg}_{\boldsymbol{\mathcal E}}\ {\rm mod}\ 2)\in \{0,1\}\times \{0,1\}\ ,
\end{align}
and we use Koszul conventions with cocycles valued in $\{\pm 1\}$ given by\footnote{An alternative convention is to assign global Grassmann parity to symbols given by the sum of fermion parity and total degree mod $2$.}
\begin{align}\label{1.9}
\boldsymbol{a}\otimes \boldsymbol{b}{}&=(-1)^{{\rm Gr}_{\boldsymbol{\mathcal E}}(\boldsymbol{a})\cdot {\rm Gr}_{\boldsymbol{\mathcal E}}(\boldsymbol{b})}\boldsymbol{b}\otimes \boldsymbol{a}\ ,\qquad \boldsymbol{a}, \boldsymbol{b}\in \boldsymbol{\mathcal E}\ ,\\
{\rm Gr}_{\boldsymbol{\mathcal E}}(\boldsymbol{a})\cdot {\rm Gr}_{\boldsymbol{\mathcal E}}{}&={\rm totfer}_{\boldsymbol{\mathcal E}}(\boldsymbol{a}){\rm totfer}_{\boldsymbol{\mathcal E}}(\boldsymbol{b})+{\rm totdeg}_{\boldsymbol{\mathcal E}}(\boldsymbol{a}){\rm totdeg}_{\boldsymbol{\mathcal E}}(\boldsymbol{b})\quad \mbox{mod $2$}\ .
\end{align}
We work with models formulated in the bosonic projections of the superbundle, viz.,
\begin{align}
\boldsymbol{\mathcal E}_0:=({\rm totfer}_{\boldsymbol{\mathcal E}})^{-1}(0)\ ,
\end{align}
with decompositions under the remaining $\mathbb{Z}$-degree, viz.,
\begin{align}
\boldsymbol{\mathcal E}_0=\bigoplus_{n\in \mathbb{Z}} \boldsymbol{\mathcal E}_{[n]}\ ,\qquad \boldsymbol{\mathcal E}_{[n]}=({\rm totdeg}_{\boldsymbol{\mathcal E}_0})^{-1}(n)\ ;
\end{align}
in particular, the superconnections, i.e., the dynamical fields, are elements of $\boldsymbol{\mathcal E}_{[1]}$.
In super-Lie algebra, we let $[\cdot,\cdot]$ denote the graded bracket.

\paragraph{$G_{\mathfrak{h}}$-tensors.}

The spin group $Spin(1,5)\cong SU^\ast(4)\equiv SL(4;\mathbb{C})\cap Sp(4;\mathbb{C})$.
We use a chirality matrix
\begin{align}
(\Gamma)_{\underline{\alpha}}{}^{\underline{\beta}}=\left[\begin{array}{cc}
\delta_\alpha^\beta&0\\0& -\delta_\beta^\alpha   
\end{array}\right]\ ,\qquad \underline\alpha,\underline\beta=1,\dots,8\ ,\qquad \alpha,\beta=1,\dots,4\ ,
\end{align}
to project 8-component Majorana spinors into 4-component left-handed (-) and right-handed (+) Weyl spinors.
The $(2,0)$-models under study have unfolded modules generated by symplectic Majorana-Weyl spinors
\begin{align}
\tau_{\underline{\alpha} I}=\left(\tau^{(+)}_{\alpha I},\tau^{(-)\alpha}{}_I\right)\ ,
\end{align}
carrying quartet-indices $I=1,\dots,4$ of the R-symmetry group $USp(4)\equiv U(4)\cap Sp(4;\mathbb{C})$.
The non-factorizable $USp(4)$-invariant tensors are $\eta_{IJ}=-\eta_{JI}$, $\eta^{IJ}=-\eta^{JI}$, and the totally anti-symmetric tensors $\epsilon_{IJKL}$ and $\epsilon^{IJKL}$, which we normalize as follows:
\begin{align}
\eta^{IJ}\eta_{IK}=\delta^J_K\ ,\qquad 
\epsilon^{IJKL}\epsilon_{I'J'K'L'}=4! \delta^{[I}_{I'}\cdots \delta^{L]}_{L'}\ ,\qquad 
\eta^{IJ}= \frac12 \epsilon^{IJKL}\eta_{KL}\ .   
\end{align}
$USp(4)$-quartet indices are raised and lowered using the conventions
\begin{align}
q^I=\eta^{IJ}q_{J}\ ,\qquad q_{I}=q^{J}\eta_{JI}\ .
\end{align}
Anti-symmetric, third rank Lorentz tensors $H_{abc}$ are decomposed into selfdual (+) and anti-selfdual (-) components according to the conventions
\begin{align}
H_{abc}=H^{(+)}_{abc}+H^{(-)}_{abc}\ ,\qquad H^{(\pm)}_{abc}:=\frac12 \left(H_{abc}\pm \frac16 \epsilon_{abcdef} H^{def}\right)=\pm\frac16 \epsilon_{abcdef} H^{(\pm)def}\ .
\end{align}
We choose chiral gamma matrices providing chiral rank-three matrices 
\begin{align}
(\sigma_{abc})_{\alpha\beta}=\frac16 \epsilon_{abcdef}(\sigma^{def})_{\alpha\beta}\ ,\qquad (\tilde\sigma_{abc})^{\alpha\beta}=-\frac16 \epsilon_{abcdef}(\tilde\sigma^{def})^{\alpha\beta}\ ,
\end{align}
which can be used to let 
\begin{align}
H^{(+)\alpha\beta}:= \frac16(\tilde\sigma^{abc})^{\alpha\beta} H^{(+)}_{abc}\ ,\qquad H^{(-)}_{\alpha\beta}:=\frac16(\sigma^{abc})_{\alpha\beta}H^{(-)}_{abc}\ .
\end{align}
Round and square brackets denote projections by symmetrization and anti-symmetrization, respectively; unaffected indices are separated using vertical bars. Young projectors ${\rm Pr}_{\vec\tau}$ are labeled by highest weights $\vec\tau$ denoted by $(\tau_1,\dots,\tau_n)$ in symmetric bases, and  $[\tau_1,\dots,\tau_n]$ in anti-symmetric bases.
Spinorial tensors are given on a symmetric basis, and R-symmetry tensors are given on an antisymmetric basis. Symmetrized groups of spinor indices and anti-symmetrized groups of $R$-symmetry indices are denoted by  $\alpha(m)\equiv (\alpha_1\dots \alpha_m)$ and $I[r]\equiv [I_1\dots I_r]$, and separated by commas, except 
\begin{align}
T_{\alpha\beta;\pm 1}=-T_{\beta\alpha;\pm 1}\ ,\qquad  N_{IJ}=N_{JI}\ ,
\end{align}
i.e., the translation and R-symmetry generators of $\mathfrak{g}_{\pm}$. 
While separate spinor indices cannot be raised or lowered, pairs of anti-symmetric spinor indices can, viz., 
\begin{align}
T^{\alpha\beta}_{\pm 1}=\frac12\epsilon^{\alpha\beta\gamma\delta}T_{\gamma\delta;\pm 1}\ .
\end{align} 
In the unfolded system, the superconformal symmetry algebra $\mathfrak{g}$ is assumed to act in modules $\boldsymbol{{\mathcal H}}_{[n]}$ of internal degree $n$ consisting of finite-dimensional irreps of the structure group $G_{\mathfrak{h}}$.
The spin-irreps are expanded using symmetric bases labelled by highest weights $(m_1,m_2,m_3)$ of $\mathfrak{su}^*(4)$ where $m_i$ are integers obeying $m_1\geqslant m_2\geqslant m_3\geqslant 0$.
The R-symmetry irreps are expanded using anti-symmetric bases labelled by highest weights $[r_1,r_2]$ of $\mathfrak{usp}(4)$ where $r_i$ are integers obeying $r_1\geqslant r_2\geqslant 0$.
Denoting the basis elements of $\boldsymbol{{\mathcal H}}_{[n]}$ by $\tau_i\equiv \tau_{[n];\Delta_i}^{(m_1,m_2,m_3);[r_1,r_2]}$, where $\Delta_i$ denotes the conformal weight defined by \eqref{3.1}, 
the corresponding fields, viz., $Z_{[1-n];\Delta_i^\ast}^{(m_1,m_2,m_3)^\ast; [r_1,r_2]^\ast}$, belong to dual representations, where the dual conformal weight
\begin{align}\label{A.26}
\Delta_i^\ast\equiv \Delta\left(Z_{[1-n];\Delta_i^\ast}^{(m_1,m_2,m_3)^\ast; [r_1,r_2]^\ast}\right):=-\Delta_i\ .
\end{align}
We introduce indices such that basis elements and fields have components $\tau_{[n];\Delta_i}^{\alpha(m_1),\beta(m_2),\gamma(m_3);I[r_1],J[r_2]}$ and $Z_{[1-n];\alpha(m_1),\beta(m_2),\gamma(m_3);I[r_1],J[r_2];\Delta^\ast_i}$, respectively.
In particular, as shown in Section \ref{zeroformmastersection}, the $T^{(-)}$ module yields component fields $Z^{(+);(m_1,m_2);[r]}_{[0];\Delta^\ast}$ with 
\begin{align}
m_1-m_2+r=2\ ,\qquad \Delta^\ast=2+\frac12(m_1+m_2)\ .
\end{align}

\section{Six-dimensional Gamma Matrices and Chiral Notation}
\label{sigmamatrices}

The Dirac matrices $\Gamma^a$, $a=0,1,\dots,5$, obey 
\begin{align}
\Gamma^a\Gamma^b+\Gamma^b\Gamma^a=2\eta^{ab}\,\qquad \eta^{ab}=(-,+,\dots,+)\ .
\end{align}
Letting $C$ denote the charge conjugation matrix, we assume that, under matrix transposition,
\begin{align}
C^T=C\ ,\qquad (C\Gamma^a)^T=- C\Gamma^a \ ,
\end{align}
i.e., we take $(t,s)=(1,5)$ and $(\epsilon,\eta)=(-1,+1)$ in the notation \cite{Sezgin:2023hkc}.
The Dirac conjugation matrix $A$ and chirality matrix $\Gamma$ are taken to be 
\begin{align}
A={}&\Gamma^0\ ,\qquad A^2=-1\ ,\qquad A^\dagger=-A\ ,\qquad 
 (CA)^T=-CA\ ,\\
\Gamma={}&\Gamma^{0}\Gamma^{1}\cdots\Gamma^{5}\ ,\qquad \Gamma^2=1\ ,\qquad \Gamma^\dagger=\Gamma\ ,\qquad (C\Gamma)^T=-C\Gamma\ , 
\end{align}
which implies
\begin{align}
({C}{A})^\ast={A}{C}^{-1}\ ,\qquad  {C}^\ast= {C}^{-1}\ .
\end{align}
We use a Weyl basis in which
\begin{align}
\Gamma_{\underline{\alpha}}{}^{\underline{\beta}} =\left[\begin{array}{cc}
    \delta_\alpha^\beta &  0\\
    0 &-\delta_\beta^\alpha 
\end{array}\right]\ ,\qquad C^{\underline{\alpha\beta}}=\left[\begin{array}{cc}
    0 &\delta^\alpha_\beta  \\
    \delta^\beta_\alpha &0 
\end{array}\right]\ ,\qquad
(\Gamma^a)_{\underline{\alpha}}{}^{\underline{\beta}} ={}&\left[\begin{array}{cc}
    0&(\sigma^a)_{\alpha\beta} \\
(\tilde{\sigma}^a)^{\alpha\beta}&0
\end{array}\right]\ ,
\end{align}
where the sigma matrices obey
\begin{align}
    (\sigma^a)_{\alpha\beta}(\tilde\sigma^b)^{\beta\gamma}+ (\sigma^b)_{\alpha\beta}(\tilde\sigma^a)^{\beta\gamma}=2\delta_\alpha^\gamma\eta^{ab},
\end{align}
and have the following symmetry and reality properties:
\begin{align}
(\tilde{\sigma}^a)^{\alpha\beta}={}&-(\tilde{\sigma}^a)^{\beta\alpha}\ ,\qquad (\sigma^a)_{\alpha\beta}=-(\sigma^a)_{\beta\alpha}\ ,\\
((\tilde{\sigma}^a)^{\alpha\beta})^\ast={}&-(\sigma^0 \tilde{\sigma}^a\sigma^0)_{\alpha\beta}\ ,\qquad ((\sigma^a)_{\alpha\beta})^\ast=-(\tilde{\sigma}^0 \sigma^a\tilde{\sigma}^0)^{\alpha\beta}\ .
\end{align}
It follows that 
\begin{align}
((\tilde{\sigma}^0)^{\alpha\beta})^\ast=(\sigma^0)_{\alpha\beta}\ ,\qquad ((\sigma^0)_{\alpha\beta})^\ast=(\tilde{\sigma}^0)^{\alpha\beta}\ ,\label{3.53}
\end{align}
and that\footnote{The Lorentz invariance implies that there exist $\lambda,\tilde \lambda\in\mathbf{C}$ such that $\epsilon^{\alpha\beta\gamma\delta}(\sigma_a)_{\gamma\delta}=\lambda(\tilde\sigma_a)^{\alpha\beta}$ and $\epsilon_{\alpha\beta\gamma\delta}(\tilde\sigma_a)^{\gamma\delta}=\tilde\lambda(\sigma_a)_{\alpha\beta}$; iteration yields $\lambda\tilde\lambda=4$. Setting $a=0$ and complex conjugating yields $\tilde \lambda = \bar\lambda$, which can also be shown by combining direct complex conjugation, which yields $\lambda^\ast = {\rm det}( \tilde\sigma^0) \lambda = ({\rm Pf} (\tilde\sigma^0))^2 \lambda$, with $\epsilon_{\alpha\beta\gamma\delta}(\tilde\sigma_a)^{\alpha\beta}(\tilde\sigma_a)^{\gamma\delta}=-4\tilde\lambda \eta^{ab}$, which yields ${\rm Pf} (\tilde\sigma^0)=\tilde\lambda/2$, and finally using $\lambda\tilde\lambda=4$. It follows that $\lambda=2\eta$ and $\tilde\lambda=2\bar\eta$.} 
\begin{align}\label{raisingloweringsigma}
\epsilon^{\alpha\beta\gamma\delta}(\sigma_a)_{\gamma\delta}=2\eta(\tilde\sigma_a)^{\alpha\beta}\ , \qquad \epsilon_{\alpha\beta\gamma\delta}(\tilde\sigma_a)^{\gamma\delta}=2\bar\eta(\sigma_a)_{\alpha\beta}\ ,
\end{align}
where $|\eta|=1$ and $\epsilon^{\alpha\beta\gamma\delta}$ and $\epsilon_{\alpha\beta\gamma\delta}$ are totally anti-symmetric tensors obeying $\epsilon^{\alpha\beta\gamma\delta}\epsilon_{\alpha\beta\gamma\delta}=4!$; we choose the overall phase of the sigma matrices such that
\begin{align}
\eta=1\ .
\end{align}
One also has the basic Fierz identity
\begin{align}
(\sigma^a)_{\alpha\beta}(\tilde\sigma_a)^{\gamma\delta}=-4\delta_{\alpha\beta}^{\gamma\delta}\ ,
\end{align}
which, for the choice of $\eta$ made above, implies 
\begin{align}
(\sigma^a)_{\alpha\beta}(\sigma_a)_{\gamma\delta}=-2\epsilon_{\alpha\beta\gamma\delta}\ , \qquad (\tilde\sigma^a)^{\alpha\beta}(\tilde\sigma_a)^{\gamma\delta}=-2\epsilon^{\alpha\beta\gamma\delta}\ .  
\end{align}
Products of pairs of sigma matrices yield two structures, viz.,
\begin{align}
(\sigma_{ab})_\alpha{}^\beta:= (\sigma_{[a})_{\alpha\gamma}(\tilde\sigma_{b]})^{\gamma\beta}\ ,\qquad (\tilde\sigma_{ab})^\alpha{}_\beta:=  (\tilde\sigma_{[a})^{\alpha\gamma}(\sigma_{b]})_{\gamma\beta}\ ,
\end{align}
related by $(\sigma_{ab})_\alpha{}^\beta=-(\tilde\sigma_{ab})^\beta{}_\alpha$.
Products of triplets of sigma matrices yield two independent structures, viz.,  
\begin{align}
(\sigma_{abc})_{\alpha\beta}=(\sigma_{[a})_{\alpha\rho_1}(\tilde\sigma_b)^{\rho_1\rho_2}(\sigma_{c]})_{\rho_2\beta}\ ,\qquad 
(\tilde\sigma_{abc})^{\alpha\beta}= (\tilde\sigma_{[a})^{\alpha\rho_1}(\sigma_b)_{\rho_1\rho_2}(\tilde\sigma_{c]})^{\rho_2\beta}\ ,
\end{align}
which are separately symmetric, i.e., $(\sigma_{abc})_{\alpha\beta}=(\sigma_{abc})_{\beta\alpha}$ and $(\tilde\sigma_{abc})^{\alpha\beta}=(\tilde\sigma_{abc})^{\beta\alpha}$.
A collection of derived Fierz identities are 
\begin{align}
(\sigma^{ab})_\alpha{}^\beta (\sigma_a)_{\gamma\delta}&=-\epsilon_{\alpha\gamma\delta\rho}(\tilde\sigma^b)^{\rho\beta}+2  (\sigma^b)^{\phantom{\beta}}_{\alpha[\gamma}\delta^\beta_{\delta]}\\
&= 4 (\sigma^b)^{\phantom{\beta}}_{\alpha[\gamma}\delta^\beta_{\delta]}+\delta_\alpha^\beta (\sigma^b)_{\gamma\delta}\ ,\\ 
(\sigma^{ab})_\alpha{}^\beta (\tilde\sigma_a)^{\gamma\delta}&=\epsilon^{\beta\gamma\delta\rho}(\sigma_b)_{\rho\alpha}-2(\tilde\sigma^b)^{\beta[\gamma}\delta_\alpha^{\delta]}\\
&=-4(\tilde\sigma^b)^{\beta[\gamma}\delta_\alpha^{\delta]}-\delta_\alpha^\beta (\tilde\sigma^b)^{\gamma\delta}\ ,\\ 
(\sigma^{ab})_\alpha{}^\beta(\sigma_{ab})_\gamma{}^\delta&= 10 \delta_{[\alpha\gamma]}^{\beta\delta}-6\delta_{(\alpha\gamma)}^{\beta\delta}=2\delta_\alpha^\beta\delta_\gamma^\delta-8\delta_\gamma^\beta\delta_\alpha^\delta\ ,\\
(\tilde\sigma^{abc})^{\alpha\beta}(\sigma_a)_{\gamma\delta}&=
-4\delta^{(\alpha}_{[\gamma}(\sigma^{bc})^{\phantom{(}}_{\delta]}{}^{\beta)}\ ,\\
(\sigma^{abc})_{\alpha\beta}(\sigma_{ab})_\gamma{}^\delta&=-8\delta^\delta_{(\alpha}(\sigma^c)_{\beta)\gamma}\ ,\\
(\tilde\sigma^{abc})^{\alpha\beta}(\sigma_{ab})_\gamma{}^\delta&=-8\delta^{(\alpha}_\gamma(\tilde\sigma^c)^{\beta)\delta}\ ,\\
(\sigma^{abc})_{\alpha\beta}(\tilde\sigma_{abc})^{\gamma\delta}&=-48\delta^{\gamma\delta}_{(\alpha\beta)}\ ,\\
(\sigma^{abc})_{\alpha\beta}(\sigma_{abc})_{\gamma\delta}&=0\ ;
\end{align}
one also has
\begin{align}
(\sigma^a)_{\alpha\rho}(\tilde\sigma_a)^{\rho\beta}={}&6\delta_\alpha^\beta\ ,
\qquad \qquad\ \qquad
(\sigma^{ab})_\alpha{}^\rho(\sigma_a)_{\rho\beta}=-5(\sigma^b)_{\alpha\beta}\ ,
\\
(\sigma^{abc})_{\alpha\rho}(\tilde\sigma_a)^{\rho\beta}={}&4(\sigma^{bc})_\alpha{}^\beta\ ,
\qquad\qquad
(\sigma^{ab})_\alpha{}^\rho (\sigma_{ab})_\rho{}^\beta=-30\delta_\alpha^\beta\ ,
\\
(\sigma^{abc})_{\alpha\rho} (\tilde\sigma_{bc})^\rho{}_\beta={}&-20 (\sigma^c)_{\alpha\beta}\ ,\qquad
(\sigma^{abc})_{\alpha\rho} (\tilde \sigma_{abc})^{\rho\beta}=-120\delta_\alpha^\beta\ ,
\end{align}
and
\begin{align}
(\sigma^{ab})_{\alpha}{}^\beta(\sigma_{cd})_{\beta}{}^{\alpha}=-8\delta^{[ab]}_{[cd]}\ ,\quad 
(\sigma^{abc})_{\alpha\beta}(\tilde\sigma_{def})^{\alpha\beta}=-24\delta^{[abc]}_{[def]}\ .    
\end{align}
Other useful identities are
\begin{align}
    &(\sigma_{[a})_{\alpha\beta}(\tilde\sigma_{b]})^{\gamma\delta}=-2
    \delta^{[\gamma}_{[\alpha}(\sigma_{ab})_{\beta]}{}^{\delta]} ,\quad 
    (\sigma_{[a})_{\alpha\beta}(\sigma_{bc]})_{\gamma}{}^{\delta}= \frac{1}{3}\delta^\delta_{[\alpha}(\sigma_{abc})_{\beta]\gamma}-\frac13\epsilon_{\alpha\beta\gamma\rho}(\tilde\sigma_{abc})^{\rho\delta} ,\\ 
    &(\tilde\sigma_{[a})^{\alpha\beta}(\tilde\sigma_{bcd]})^{\gamma\delta}=-\frac{1}{4} \epsilon^{\alpha\beta(\gamma|\rho}(\sigma_{abcd})_{\rho}{}^{|\delta)}\ .
\end{align}
Similar ones can also be found using \eqref{raisingloweringsigma}.

\section{Linearized Unfolded Systems on Commutative Sources}
\label{App:symmetriescocycles}

This Appendix describes linearized, unfolded systems on commutative sources.
The first part describes systems consisting of connection one-forms in Lie (super)algebras and spaces of forms in tensorial modules glued via cocycles.
The second parts describe the linearization of nonlinear unfolded systems based on homotopy Lie algebras around backgrounds.

\subsection{Ab initio linearized system}

\paragraph{Cocycles.}

The restriction of the graded symmetry group of an unfolded system to degree zero is a Lie (super)group $G$ with Lie (super)algebra $\mathfrak{g}$.
We assume that the structure group of the system is a sub(super)group $G_{\mathfrak{h}}$ of $G$ with Lie (super)algebra $\mathfrak{h}$.
Various boundary conditions of the system are encoded into (possibly projective) irreducible $G$-modules  $\mathsf{W}_\lambda$; we use conventions in which $G$ acts on $\mathsf{W}_\lambda$ from the left.
To provide manifestly $G_{\mathfrak{h}}$-invariant couplings between these modules, one assumes injective $\mathfrak{g}$-morphisms 
\begin{align}
\tau:\mathsf{W}_\lambda\downarrow_{\mathfrak{g}}\to \mathsf{T}_\lambda\ ,
\end{align}
to $\mathfrak{g}$-irreps\footnote{In the infinite-dimensional case, $\mathsf{T}_\lambda$ may fail to be a $G$-module, in which case classical singularities may arise in the component fields dual to $\tau_{\lambda;s}$ \cite{families,BTZ2019,corfu21}.} $\mathsf{T}_\lambda$ on which $G_{\mathfrak{h}}$ acts faithfully, i.e.,
\begin{align}
\mathsf{T}_\lambda\downarrow_H=\bigoplus_s{}^{\!{}^\sigma} \tau_{\lambda;s}  
\end{align}
is a semi-direct sum of $G_{\mathfrak{h}}$-irreps $\tau_{\lambda;s}$ glued via $G_{\mathfrak{h}}$-equivariant cocycles $\sigma_s^{s'}:\tau_{\lambda;s'}\to \tau_{\lambda;s}$; for simple $G_{\mathfrak{h}}$ and finite-dimensional\footnote{For examples of unfolded, relativistic field theories with infinite-dimensional, fractional-spin representations of the Lorentz group, see \cite{Boulanger:2013naa,Boulanger:2015uha}.
We expect non-trivial cocycles to arise in unfolded formulations of Carrollian field theories with non-semi-simple structure groups given by contractions of the Lorentz group.
} 
$\tau_{\lambda;s}$, all $\sigma_s^{s'}=0$, i.e., 
\begin{align}
\mathsf{T}_\lambda\downarrow_H=\bigoplus_s \tau_{\lambda;s}
\end{align} 
is a direct sum of $G_{\mathfrak{h}}$-irreps.
Letting $\boldsymbol\Omega \in \Omega_{[1]}(\boldsymbol{M})\otimes \mathfrak{g}$ be a classical vacuum solution, viz.,
\begin{align}
d\boldsymbol\Omega+\boldsymbol\Omega\wedge \boldsymbol\Omega\approx 0\ ,
\end{align}
linearized fluctuations around it form a master field 
\begin{align}
\boldsymbol{x}=\bigoplus_\lambda{}^{\!{}^\Sigma}\,\boldsymbol{x}_\lambda\ ,\qquad \boldsymbol{x}_\lambda\in \Omega_{[p_\lambda]}(\boldsymbol{M})\otimes\mathsf{W}_\lambda    \ ,
\end{align} 
obeying
\begin{align}\label{C.6}
d\boldsymbol\Omega+\boldsymbol\Omega\wedge \boldsymbol\Omega\approx 0\ ,\qquad 
d\tau(\boldsymbol{x})+ \boldsymbol\Omega \tau(\boldsymbol{x})+\boldsymbol{\Sigma}(\tau(\boldsymbol{x}))\approx 0\ ,
\end{align}
where 
\begin{align}
\boldsymbol{\Sigma}=\bigoplus_{\lambda;\lambda'}\boldsymbol{\Sigma}_\lambda^{\lambda'}\ ,\qquad \boldsymbol{\Sigma}_\lambda^{\lambda'}:\Omega_{[p_{\lambda'}]}(\boldsymbol{M})\otimes\mathsf{T}_{\lambda'}\to \Omega_{[p_\lambda]}(\boldsymbol{M})\otimes\mathsf{T}_\lambda\ ,
\end{align}
that is, $\boldsymbol{\Sigma}_\lambda^{\lambda'}\in \Omega_{[1+p_{\lambda}-p_{\lambda'}]}(\boldsymbol{M})\otimes {\rm Hom}(\mathsf{T}_{\lambda'},\mathsf{T}_\lambda)$, are $G_{\mathfrak{h}}$-equivariant cocycles compatible with universal Cartan integrability.
Thus, embedding a connection one-form $\boldsymbol{\omega}$ of the principal $G_{\mathfrak{h}}$-bundle over $\boldsymbol{M}$ into the background connection via a non-canonical projection ${\rm pr}: \mathfrak{g}\to \mathfrak{g}$, viz.,
\begin{align}
\boldsymbol{\omega}={\rm pr}(\boldsymbol{\Omega})\ ,
\end{align}
and letting $\boldsymbol{E}:=(1-{\rm pr})(\boldsymbol{\Omega})$ and $\boldsymbol{\nabla}:=d+\boldsymbol{\omega}$, 
the system takes the form
\begin{align}
\boldsymbol{\nabla}\boldsymbol{E}+(1-{\rm pr})(\boldsymbol{E}\wedge \boldsymbol{E}){}&\approx0\ ,\qquad \boldsymbol{\nabla}^2+{\rm pr}(\boldsymbol{E}\wedge \boldsymbol{E})\approx 0\ ,\\
\label{B.6}
\boldsymbol{\nabla} \tau(\boldsymbol{x}_\lambda)+\boldsymbol{E}\tau(\boldsymbol{x}_\lambda)+\sum_{\lambda'}\boldsymbol{\Sigma}_{\lambda}^{\lambda'}(\tau(\boldsymbol{x}_{\lambda'})){}&\approx 0\ ,
\end{align}
where $\boldsymbol{\Sigma}_\lambda^{\lambda'}$ are functions of $\boldsymbol{E}$ obeying 
\begin{align}
[\boldsymbol{E},\boldsymbol{\Sigma}_\lambda^{\lambda'}]-(1-{\rm pr})(\boldsymbol{E}\wedge \boldsymbol{E}) \cdot \frac{\partial}{\partial \boldsymbol{E}} \Sigma_\lambda^{\lambda'}-\sum_{\lambda''} (-1)^{p_\lambda-p_{\lambda''}}\Sigma_\lambda^{\lambda''}\Sigma_{\lambda''}^{\lambda'}\equiv 0\ ,
\end{align}
modulo shifts by trivial solutions, i.e., the cocycles belong to a graded Chevalley--Eilenberg cohomology group \cite{Vasiliev:1988sa,Vasiliev:1999ba,Bekaert:2004qos,BMVI,Sharapov:2017yde,Sharapov:2020quq}. 

\paragraph{Weyl zero-form module.}

On-shell, $\boldsymbol{x}$ consists of homogeneous, i.e., pure gauge, solutions in strictly positive degrees and particular solutions sourced by the integration constants in degree zero via the cocycles.
Constant, i.e., $\boldsymbol{E}$-independent, cocycles correspond to contractible pairs, i.e., spontaneously broken gauge symmetries.
In an unfolded relativistic field theory, the local degrees of freedom arise as integration constants of the submodule of the zero-form module remaining after all contractible zero-forms have been removed.
In the context of extensions of ordinary gravity, the contracted zero-form module is referred to as the Weyl zero-form module \cite{Vasiliev:1999ba,Bekaert:2004qos,Didenko:2014dwa}; in the context of six-dimensional models with chiral curvatures on-shell, we refer to them as chiral zero-form module.

\paragraph{Gauge symmetries.}

The universal Cartan integrability implies two types of gauge transformations, viz.,
\begin{align}
\delta_{\boldsymbol{\Lambda};\boldsymbol{\epsilon}} \boldsymbol{\Omega}&=d\boldsymbol{\Lambda}+[\boldsymbol{\Omega},\boldsymbol{\Lambda}]\ ,\\
\delta_{\boldsymbol{\Lambda};\boldsymbol{\epsilon}} \tau(\boldsymbol{x}_\lambda)&=d\tau(\boldsymbol{\epsilon}_\lambda)+ \boldsymbol\Omega \tau(\boldsymbol{\epsilon}_\lambda)+\sum_{\lambda'}(-1)^{p_\lambda-p_{\lambda'}}\boldsymbol{\Sigma}(\tau(\boldsymbol{\epsilon}_{_\lambda'}))\\
&-\boldsymbol{\Lambda} \tau(\boldsymbol{x}_\lambda)- \sum_{\lambda'}(1-{\rm pr})(\boldsymbol{\Lambda})\cdot \left(\frac{\partial}{\partial \boldsymbol{E}} \boldsymbol{\Sigma}_\lambda^{\lambda'} \right)(\tau(\boldsymbol{x}_{\lambda'}))\ ;
\end{align}
where $\delta_{\boldsymbol{\Lambda};0}$ are non-abelian symmetries with $\mathfrak{g}$-valued parameters $\boldsymbol{\Lambda}$ associated with $\boldsymbol{\Omega}$, and $\delta_{0;\boldsymbol{\epsilon}}$ are abelian symmetries with $\mathsf{W}_\lambda$-valued parameters $\boldsymbol{\epsilon}_\lambda$ associated with $\boldsymbol{x}$.
The former contain background symmetries with rigid parameters $\overline{\boldsymbol{\Lambda}}$, i.e.,
\begin{align}
\delta_{\overline{\boldsymbol{\Lambda}};0} \Omega=0\ ,\qquad \delta_{\overline{\boldsymbol{\Lambda}};0} \tau(\boldsymbol{x})=-\overline{\boldsymbol{\Lambda}} \tau(\boldsymbol{x})- (1-{\rm pr})(\overline{\boldsymbol{\Lambda}})\cdot \left(\frac{\partial}{\partial \boldsymbol{E}}\Sigma \right)(\tau(\boldsymbol{x}))\ ,
\end{align}
acting via representation matrices in form-degree zero and differentiated cocycles 
in positive degrees.

\subsection{Linearization of nonlinear system}

\paragraph{Nonlinear systems.}

Starting from \eqref{2.36}, we expand
\begin{align}\label{C.16}
\check{\boldsymbol{z}}=\check{Z}^i\otimes {\tau}_i\ ,\qquad \sum_{r=1}^\infty  \check{l}_r(\check{\boldsymbol{z}}^{\wedge r})\equiv \check{Q}^i(\check Z)\otimes {\tau}_i\ ,
\end{align}
where $\check Z^i\in \Omega_{[0]}(\mathcal{M})\otimes \Omega(\check{\boldsymbol{M}})$, and the generating functions obey
\begin{align}
\check{Q}^i \wedge \check{\partial}_i \check{Q}^j\equiv 0\ ;
\end{align}
the resulting component form of the equations of motion read
\begin{align}\label{C.17}
\check R^i:= d_{\check{\boldsymbol{M}}}\check{Z}^i- \check{Q}^i\approx 0\ ,
\end{align}
where the Cartan curvatures obey Bianchi identities, viz.,
\begin{align}
d\check{R}^i+\check{R}^j \wedge \check{Q}_j{}^i\equiv 0\ ,\qquad \check{Q}_j{}^i:=\check{\partial}_j \check{Q}^i\ ,
\end{align}
that is, \eqref{C.17} are integrable universally, i.e., on any graded commutative source $\check{\boldsymbol{M}}$.
Expanding $\check{\boldsymbol{\epsilon}}=\check{\epsilon}^i \otimes {\tau}_i$, where $\check{\epsilon}^i\in \Omega_{[0]}(\mathcal{M})\otimes \Omega(\check{\boldsymbol{M}})$, the gauge transformations 
\begin{align}
\delta_{\check{\boldsymbol{\epsilon}}} \check{Z}^i=\check{T}^i_{\check{\boldsymbol{\epsilon}}} :=d\check{\epsilon}^i+ \check{\epsilon}^j \wedge \check{Q}_j{}^i\ ,
\end{align}
act faithfully on the shell, viz.,
\begin{align}
\delta_{\check\epsilon} \check{R}^i=(-1)^j \check{\epsilon}^j \wedge\check{R}^k \wedge\check{Q}_{kj}{}^i\ ,\qquad \check{Q}_{kj}{}^i:=\check{\partial}_k\check{\partial}_j \check{Q}^i\ ,
\end{align}
and close on-shell, viz.,
\begin{align}
[\delta_{\check\epsilon_1},\delta_{\check\epsilon_2}]\check{Z}^i=\delta_{\check\epsilon_{12}}\check{Z}^i+  \check{\epsilon}^j_1 \wedge\check{\epsilon}^k_2 \wedge\check{R}^l \wedge\check{\partial}_l\check{\partial}_k\check{\partial}_j \check{Q}^i\ ,\qquad \check{\epsilon}^i_{12}:=\check{\epsilon}^j_1 \wedge\check{\epsilon}^k_2 \wedge\check{Q}_{kj}{}^i\ ,
\end{align} 
with off-shell closure obstructed by the higher brackets of $\check{\boldsymbol{\mathcal L}}$, i.e., $\check{l}_r$, $r\geqslant 3$.
Locally, the classical solution space decomposes into Cartan gauge orbits consisting of solutions of the form
\begin{align}\label{C.23}
\check{Z}^i_{\check{\lambda};\check{\boldsymbol{c}}}=e^{\check{T}^j_{\check{\boldsymbol{\lambda}}}\check{\partial}_j} \check{Z}^i|_{\check{\boldsymbol{z}}=\check{\boldsymbol{c}}}\ ,
\end{align}
where $\check{\boldsymbol{c}}$ consists of zero-form integration constants $\check{C}^i$, i.e., $\check{C}^i$ is a constant if ${\rm deg}(\check{C}^i)=0$ and vanishes else.

\paragraph{Expansion around background.}

Assuming that $\overline{Z}^i$ is a classical background,  splitting
\begin{align}
\check{Z}^i=\overline{Z}{}^i+ \stackrel{(1)}{Z}{}^{\!\!i}\ ,
\end{align}
and expanding 
\begin{align}
\check{R}^i\equiv \overline{R}^i + \stackrel{(1)}{R}{}^{\!\!i}+O\left(\left(\stackrel{(1)}{Z}\right)^2\right)\ ,\qquad \stackrel{(1)}{R}{}^{\!\!i} := d_{\check{\boldsymbol{M}}}\stackrel{(1)}{Z}{}^{\!\!i}-\stackrel{(1)}{Z}{}^{\!\!j}\wedge \overline{Q}_j{}^i\ ,
\end{align}
it follows that the linearized curvatures obey the Bianchi identities
\begin{align}
d_{\check{\boldsymbol{M}}}\stackrel{(1)}{R}{}^{\!\!i}+\stackrel{(1)}{R}{}^{\!\!j}\wedge \overline{Q}_j{}^i\equiv 0\ ,
\end{align}
such that the linearized fluctuations around the background are captured by the universally Cartan integrable system
\begin{align}
\overline{R}^i\approx0\ ,\qquad \stackrel{(1)}{R}{}^{\!\!i}\approx0\ ,
\end{align}
with $\overline{Z}{}^i$ and $\stackrel{(1)}{Z}{}^{\!\!i}$ treated as independent fields, and corresponding gauge symmetries
\begin{align}
\delta_{\boldsymbol{\overline\Lambda};\stackrel{(1)}{\boldsymbol{\epsilon}}} \overline Z{}^i{}&=d_{\check{\boldsymbol{M}}}\overline\Lambda{}^i+\overline\Lambda{}^j\wedge \overline{Q}_j{}^i\ ,\\
\delta_{\boldsymbol{\overline\Lambda};\stackrel{(1)}{\boldsymbol{\epsilon}}}  \stackrel{(1)}{Z}{}^{\!\!i}{}&=d_{\check{\boldsymbol{M}}}\!\stackrel{(1)}{\epsilon}{}^{\!\!i}+\stackrel{(1)}{\epsilon}{}^{\!\!j}\wedge \overline{Q}_j{}^i+(-1)^j \stackrel{(1)}{Z}{}^{\!\!j}\wedge \overline{\Lambda}{}^k\wedge \overline{Q}_{kj}{}^i\ ,
\end{align}
where $\delta_{0;\stackrel{(1)}{\boldsymbol{\epsilon}}}$ are nilpotent (hence abelian).
Background symmetries correspond to parameters $\overline{\Lambda}{}^i$ obeying 
\begin{align}\label{C.30}
\delta_{\overline{\boldsymbol{\Lambda}};0}\overline{Z}{}^i=0\ ,
\end{align}
which together with $\overline{R}{}^i\approx 0$ constitute a universally Cartan integrable system with zero-form constraints, i.e., \eqref{C.30} decompose under form-degree on $\check{\boldsymbol{M}}$ into algebraic constraints in degree zero and constraints on $d_{\check{\boldsymbol{M}}}\overline{\Lambda}{}^i$ in strictly positive degree that are integrable universally modulo the former.
In particular, if $\overline{Z}^i$ vanishes in form-degree zero and $\check{l}_1(\check{\boldsymbol{\mathcal L}}_{[0]})=0$, i.e., in the absence of Stuckelberg zero-forms, then $\delta_{\overline\lambda;0} \overline X^A$ vanishes in form-degree zero, i.e., \eqref{C.30} does not contain any zero-form constraints, and the background symmetry parameters represent $\check{\boldsymbol{\mathcal L}}$.
Restricting $\overline{Z}^i$ further to form-degree one yields a system of the type introduced in \eqref{C.6}.

\end{appendix}

\providecommand{\href}[2]{#2}\begingroup\raggedright\endgroup

\end{document}